\documentclass[a4paper,oneside,reqno,12pt]{article}
\usepackage{amsmath,amsthm,amssymb,epsfig,latexsym}
\usepackage[pdftex]{hyperref}
\usepackage{vmargin}        % Rgler la taille de la feuille.
\setmarginsrb{1.8cm}{1cm}{1.8cm}{2cm}{1cm}{1cm}{1cm}{1.6cm}
\def\Cbbd{\mathbb{C}}

\def\Rbbd{\mathbb{R}}

\def\Zbbd{\mathbb{Z}}

\def\Ical{\mathcal{I}}
\def\Jcal{\mathcal{J}}

\def\Vcal{\mathcal{V}}

\def\Ibf{\boldsymbol{I}}

\def\vbf{\boldsymbol{v}}

\def\gbf{\boldsymbol{\gamma}}

\newcommand{\rd}{\mathrm{d}}            % Roman d for differential
\newcommand{\re}{\mathrm{e}}            % Roman e for exponential
\newcommand{\ri}{\mathrm{i}}            % Roman i for imaginary number
\def\a{\alpha}

\def\g{\gamma}
\def\d{\delta}
\def\eps{\varepsilon}
\def\la{\lambda}
\def\s{\sigma}
\def\phi{\varphi}

\def\D{\Delta}

\def\Im{\mathop{\hbox{\rm Im}}\nolimits}
\def\Re{\mathop{\hbox{\rm Re}}\nolimits}

\def\abs#1{\left|#1\right|}
\def\dd{\partial}
\def\Ref#1{(\ref{#1})}

\def\endproof{\hfill\rule{2mm}{2mm}}
\def\id{\text{id}}
\def\endproof{\hfill\rule{2mm}{2mm}}
\def\?{(?)\marginpar[\hfill?|]{|?}}

\def\matmz{\mathop{\hbox{\rm Mat}}\nolimits_M(\Zbbd)}

\newcommand{\stirling}[2]{\genfrac{[}{]}{0pt}{}#1#2}
\def\beq{\begin{equation}}
\def\eeq{\end{equation}}
\def\be{\begin{equation*}}
\def\ee{\end{equation*}}

\newtheorem{theo}{Theorem}
\newtheorem{prop}{Proposition}
\newtheorem{lemme}{Lemma}
\newtheorem{defin}{Definition}
\newtheorem{cor}{Corollary}

 \makeatletter
 \@addtoreset{equation}{section}
 \makeatother

\begin{document}

%\begin{flushright}
%YorkU-2009/2010-12
%\end{flushright}
%\par \vskip .1in \noindent
%
%\vspace{14pt}

\begin{center}
\begin{LARGE}
{\bf Combinatorics of generalized Bethe equations}
\end{LARGE}

\vspace{30pt}

\begin{large}
{\bf K.~K.~Kozlowski\footnote[1]{Universit\'{e} de Bourgogne, Institut  de Math\'{e}matiques de Bourgogne, UMR 5584 du CNRS,  France, karol.kozlowski@u-bourgogne.fr}  
and
E.~K.~Sklyanin}\footnote[2]{ Department of Mathematics, University of York, York YO10 5DD, UK,
eks2@york.ac.uk;
work supported by EPSRC grant EP/H000054/1}\\
\par

\end{large}

\vspace{40pt}

\centerline{\bf Abstract} \vspace{1cm}
\parbox{12cm}{\small  A generalization of the Bethe ansatz equations is studied,
where a scalar two-particle S-matrix has several zeroes and poles in the complex plane,
as opposed to the ordinary single pole/zero case.
For the repulsive case (no complex roots), the main result is the enumeration of all distinct solutions 
to the Bethe equations in terms of the Fuss-Catalan numbers. Two new combinatorial interpretations of the Fuss-Catalan and related numbers are obtained. On the one hand, they count regular orbits of the permutation group in certain factor modules over $\mathbb{Z}^M$, and on the other hand, they count integer points in certain $M$-dimensional polytopes.
}
\end{center}

\vspace{40pt}

\newpage

\section*{Introduction}

The modern techniques of quantum integrability rely on what is known today as the {\em Bethe ansatz},
a powerful tool, first introduced by H.~Bethe \cite{BetheSolutionToXXX} and refined by many authors
throughout the years, see the summarising monographs \cite{BaxterExactlySolvableLatticeModels,BogoliubiovIzerginKorepinBookCorrFctAndABA,FaddeevABALesHouchesLectures}.
The idea behind the Bethe ansatz is to reduce, in one way or another, the spectral problem for the
Hamiltonian in the $M$-particle sector to the one of solving a set of $M$ simultaneous equations (algebraic or transcendental, depending on the model) for particles' {\em rapidities} $v_n$.

In the case when the particles carry no internal degrees of freedom (are spinless), those equations, referred to as {\em Bethe equations}, can be cast in the form
\beq\label{bethe}
    F(v_n)\prod_{\substack{m=1\\ m\neq n}}^M S(v_n-v_m)=1, \qquad
    n=1,\ldots,M,
\eeq
and interpreted as the periodicity conditions for the wave function of the
system of $M$ one-dimensional particles put in a periodic box, in the assumption that
all multi-particle collisions can be reduced to a sequence of two-particle collisions
({\em factorisation of $S$-matrix} hypothesis).
The function $F$ can be interpreted then as a one-particle travel phase factor, and $S$ as a two-particle scattering phase factor. 

Most of the solutions of the equations \Ref{bethe}, referred to in the present paper as 
{\em weak} generalized Bethe equations (weak GBE), correspond to the discrete eigenstates
of the integrable system. To eliminate the nonphysical solutions one needs, 
as explained in section \ref{Section Generalites}, to supplement \Ref{bethe} with additional
conditions. Such corrected GBE will be called {\em strong} GBE.

Although  $F$ can take a wide range of forms depending on the chosen integrable model, the expression for $S$ is much less variable,
being restricted by the quantum group governing the integrability of the model.
In the most of the presently known examples,
the $S$-factor has only a single pole and a single zero in the complex plane (modulo periodicity for trigonometric or elliptic functions).
The only exception known so far is the Izergin-Korepin model \cite{ReshetikhinMethodOfFunctionalEquations,TarasovABAforIzerginKorepin}.

In the present paper we start a systematic study of the Bethe equations,
where the function $S$ has $K$ poles and $K$ zeroes. 
Our primary interest is the case $K>1$, referred to as the {\em generalized} Bethe equations
(GBE), as opposed to the well-studied {\em ordinary} case $K=1$.
Our main motivation comes from the recent 
progress on AdS/CFT correspondence in string theory that produced a similar kind of Bethe equations
\cite{KazakovMarshakovMinahanZaremboIntegrabilityInAdS,BargheerBeisertLoebbertLongRangeDeformations}.
It has to be emphasized that in contrast to the ``asymptotic Bethe ansatz'' approach 
as in \cite{KazakovMarshakovMinahanZaremboIntegrabilityInAdS,BargheerBeisertLoebbertLongRangeDeformations},
where the equations \eqref{bethe} hold in the infinite volume limit only, we assume here that they are satisfied exactly in the finite volume as well.

We shall concentrate on the simplest task: counting the number of solutions to the strong Bethe equations.
Even in the ordinary case of a single pole/zero $S$-function, the combinatorics of Bethe ansatz
related to the so-called ``string hypothesis'' and ``completeness problem''
has been found to be rewardingly rich and interesting
\cite{FaddeevTakhtadzhanXXXgeneralOverwiev,KirillovReshetikhinYangiansBetheAnsatzCombinatorics}.
Mathematically, the task is to count solutions of a special system of algebraic equations.
We hope that our results present interest not only for the specialists in integrable systems
but also in combinatorics.

To avoid complications caused by complex solutions we focus on the {\em repulsive} case, when
all roots $v_n$ are real. In this case, the {\em strong} conditions are equivalent simply to discarding
multiple roots of the Bethe equations.
We consider two cases: the functions $F$ and $S$ being ratios of trigonometric or ordinary polynomials of degree $N$ for $F$ and degree $K$ for $S$. The number of distinct solutions to the strong GBE is then denoted $C^{(K,N)}_M$.

To count the solutions of GBE we use the fact that they correspond to the minima of certain convex function (Yang-Yang function).
In the trigonometric case, we establish a one-to-one correspondence between the set of distinct solutions to the GBE and the set of regular orbits of the permutation group $\mathfrak{S}_M$ in certain factor modules over $\Zbbd^M$. In the rational case, we bijectively label the solutions with the integer sequences $I_1<\dots<I_M$ satisfying certain linear inequalities. For a more detailed description see respectively theorems 1 and 2 below.

For the numbers $C^{(K,N)}_M$ that enumerate both combinatorial objects and therefore the solutions of strong GBE as well, we then obtain an explicit formula
\beq\label{cmkn}
   C_M^{(K,N)}=\frac{N}{M}\,\binom{KM+N-1}{M-1} 
              =\frac{N}{KM+N}\,\binom{KM+N}{M}
\eeq
that constitues the main result of our paper.

The respective generating function turns out to be the $N^{\text{th}}$ power of the generating function for the Fuss-Catalan numbers \cite{GrahamKnuthPatashnikConcreteMathematics,StanleyEnumerativeCombinatorics} corresponding to $C^{(K,1)}_M$ in our notation:
\begin{equation}
   C^{(K,N)}(t)= \sum_{M=0}^{\infty} C_M^{(K,N)} t^{M} = \biggl(C^{(K,1)}(t)\biggl)^N.
%\left(\sum_{M=0}^{\infty} C_M^{(K,1)} t^{M}\right)^N.
\end{equation}

The particular case $C_M^{(2,1)}$ corresponds to the famous
Catalan numbers \cite{HiltonPedersen} that are known to have over 120 different combinatorial interpretations \cite{StanleyCatalanAddendum}.

Two theorems formulated below provide two new combinatorial interpretations of the numbers $C^{(K,N)}_M$, including Fuss-Catalan numbers $C^{(K,1)}_M$ and Catalan numbers $C^{(2,1)}_M$ as particular cases,
obtained as a result of our study of GBE. 

\begin{theo}
\label{theorem comptage elements factor module}
Let $N$, $M$ and $K$ be some positive integers.  Two $M$-tuples of integers $(I_1,\dots I_M)$ are considered equivalent if one can be mapped into another by a finite composition of basic transformations $\gamma_n$ ($n=1,\ldots,M$)
\beq\label{definition operation shift}
     \gamma_n: I_n \mapsto I_n + N+(M-1)K  \qquad \text{and} \qquad 
     \gamma_n: I_m \mapsto I_m -K  \quad m\not= n \; ,
\eeq

An equivalence class is called {\em regular} if it contains only $M$-tuples of pairwise distinct
integers $(I_1,\dots I_M)$.

Then the number of all regular equivalence classes is given by $M!\,C_M^{(K,N)}$.
\end{theo}

\begin{theo}
\label{theorem integers with constraints}
Let $N$, $M$ and $K$ be some positive integers, then the number  of different choices of strictly increasing integer sequences $I_1<\dots<I_M$ satisfying to $2M$ constraints
\begin{subequations}
\beq
   \frac{r(r-1)}{2}\, K \,\,\leq\,\,  I_1+I_2+\ldots+I_r,
\eeq
\beq
    I_{M-r+1}+\ldots+I_{M-1}+I_M \,\,\leq\,\,  r M K  -\frac{r(r+1)}{2} K + r N -1,
\eeq
\end{subequations}
for $r=1,\dots , M$ equals $C^{(K,N)}_M$.
\end{theo}

The paper is organized as follows. In Section \ref{Section Generalites}, we give a precise description of the class of functions $F$ and $S$ that we consider. We then define the strong generalized Bethe equations (GBE)
and prove certain elementary properties of the solutions to
the weak and strong GBE associated with our choice of $F$ and $S$. In Section \ref{Section criteres de solvabilite}, we reformulate GBE as the minimality condition for a convex Yang-Yang function and establish a one-to-one correspondence between the set of distinct solutions to the strong GBE and the sets of integers described in theorems \ref{theorem comptage elements factor module} and \ref{theorem integers with constraints}. Then, we distinguish between the trigonometric and rational cases. In Section \ref{Section denombrement solutions GBAE}, we apply the theory of commutative rings to count the distinct solutions to the strong GBE in the trigonometric case and prove theorem \ref{theorem comptage elements factor module}. Finally, in Section \ref{section preuve theroem comptage cas rationel}, we analyze the limit when the period of trigonometric functions goes to infinity
and show that the number of solutions in the rational case is the same as in the trigonometric case.
Theorem~\ref{theorem integers with constraints} is then obtained as a corollary of the established bijection.

The important question as to what kind of integrable models 
and what quantum groups might correspond to our generalized Bethe equations 
is left unresolved. We plan
to address it in our subsequent publications.

%%%%%%%%%%%%%%%%%%%%%%%%%%%%%%%%%%%%%%%%%%%%%%%%%%%%%%%%%%%%%%%%%%%%%%%%%%%%%%%%%%%%%%%%%%%%%%%%%%%%%%
%%%%%%%%%%%%%%%%%%%%%%%%%%%%%%%%%%%%%%%%%%%%%%%%%%%%%%%%%%%%%%%%%%%%%%%%%%%%%%%%%%%%%%%%%%%%%%%%%%%%
\section{The general setting}
\label{Section Generalites}

Throughout the paper, we consider two cases of interest. We assume that $F$ and $S$
are either realized as ratios of ordinary polynomials or as ratios of trigonometric polynomials. We will refer to the first case as the \textit{rational} case and to the second one as the  \textit{trigonometric} case.

Let $\psi(u)=u$ in the  rational case and $\psi(u)=\sin(u)$ in the trigonometric one. The functions
$F$ and $S$ are characterized by natural numbers $N, K$ and complex parameters
$F_\infty, \xi_n, \eta_j\in\Cbbd$ and are defined as 
\begin{subequations}\label{defFS}
\begin{alignat}{3}
    F(u)&=F_\infty\,\prod_{n=1}^N \frac{\psi(u+\xi_n)}{\psi(u+\overline{\xi}_n)}\,,&\qquad
    \abs{F_\infty}&=1, &\qquad & \Im(\xi_n)>0\,, \; \; n \in \{1,\dots,N\},
\label{definition fonction F} \\
    S(u)&= \prod_{j=1}^{K} \frac{\psi(u+\eta_j)}{\psi(u+\overline{\eta}_j)}\,,
    &\qquad&
    &\qquad &\Im(\eta_j)>0\,,
     \; \; j\in \{1,\dots,K\} \; .
\label{definition fonction S}
\end{alignat}
\end{subequations}

Here $\overline{z}$ stands for the complex conjugate of $z$.
The parameters $\eta_j$ are chosen in such a way that the sets of zeroes and poles of $S$ are both symmetric
with respect to the imaginary axis: 
\beq\label{symmetry condition on etas}
   \{\eta\}=\{-\overline\eta\,\}.
\eeq

To avoid degeneracy of solutions to GBE, we demand that $F_{\infty}\neq1$, or, equivalently,  $F_{\infty}=\text{e}^{-2i\pi\varphi_{\infty}}$, with $-1< \varphi_{\infty}<0$.
Later, in some cases we impose more restrictive conditions on $\varphi_{\infty}$.

Finally, in the trigonometric case, we also assume that the parameters $\xi_n$ and $\eta_j$ are located
in the strip $-\pi/2<\Re(z)\leq\pi/2$.

The above restrictions on $F$ and $S$ are chosen in such a way that the phase factors, both trigonometric and rational, have the following natural, for a physicist, properties that are easy to derive from \eqref{defFS} and \eqref{symmetry condition on etas}.

\medskip
{\bf 1. Unitarity.}
\beq
    \overline{F(u)}=F(\overline{u})^{-1}, \quad \overline{S(u)}=S(\overline{u})^{-1},
    \qquad u\in\Cbbd.
\label{unitarity}
\eeq

{\bf 2. Parity.} $S$ respects the space-, or $P$-parity:
\begin{subequations}
\beq
    S(-u)=S(u)^{-1},
\label{P-parity}
\eeq
and the time-, or $T$-parity:
\beq
   \overline{S(u)}=S(-\overline{u}).
\label{T-parity}
\eeq
\end{subequations}

{\bf 3. Repulsion condition:}
\begin{subequations}\label{repulsion_cond}
\begin{alignat}{3}
  \abs{F(u)}&>1, &\quad \abs{S(u)}&>1, &\quad \text{\rm }\quad \Im u&>0 \, , \\
  \abs{F(u)}&<1, &\quad \abs{S(u)}&<1, &\quad \text{\rm }\quad \Im u&<0 \, .
\end{alignat}
\end{subequations}

{\bf Remark.} The unitarity condition implies that
 $   \abs{F(u)}=1=\abs{S(u)}$ for $u\in\Rbbd $.
Also it is easy to see that the two parity conditions \Ref{P-parity} and \Ref{T-parity} are equivalent as long as one assumes unitarity.

It is a well known fact in the literature on Bethe ansatz that the Bethe equations \eqref{bethe}, taken literally as they are written, have too many ``false'' solutions that do not correspond to any physical states. A common trick to eliminate them from consideration is to replace \eqref{bethe} with a different, more restrictive problem for $v_m$ in terms of the Baxter polynomial:
\beq\label{defQ}
   Q(u|\,\boldsymbol{v})\equiv\prod_{m=1}^M \psi(u-v_m),\qquad
   \boldsymbol{ v } =(v_1,\dots, v_M).
\eeq

The problem for $Q(u|\,\boldsymbol{v})$ described below generalizes Baxter's $TQ$ equation known in the ordinary $K=1$ case \cite{BaxterExactlySolvableLatticeModels}. An important difference is that for $K>1$ the equation for $Q$ is not linear.

Let ${  \bf z } =(z_1,\dots,z_M)\in\Cbbd^M$ and 
$\mathcal{Q}_{\{\eta\},\{\xi\}}(u|\,{\bf z})$ be defined as
\beq
\mathcal{Q}_{\{ \eta\},\{ \xi\}}(u|\,{\bf z}) 
  = \alpha(\mu|\,\{ \xi\})
  \prod_{k=1}^{K} Q(u+\eta_k|\,\boldsymbol{z})
\; + \; \delta(u|\,\{\xi\}) \prod_{k=1}^{K} Q(u+\overline{\eta}_k |\,\boldsymbol{z}) \; .
\label{definition Q}
\eeq
with
\beq
\alpha(u|\,\{\xi\})= F_{\infty} \prod_{n=1}^{N} \psi(u+\xi_n) \; ,
\qquad \delta (u|\,\{\xi\}) = (-1)^{K+1}\prod_{n=1}^{N} \psi(u+\overline{\xi}_n) \; .
\eeq

We define the {\em strong} GBE
to be the set of $M$ conditions on the coordinates of the vector
$\boldsymbol{ v } =(v_1,\dots, v_m)$ necessary and sufficient that the function
\beq\label{QoverQ}
u \mapsto \frac{ \mathcal{Q}_{\{ \eta\},\{ \xi\}}(u|\,\boldsymbol{v}) }{Q(u|\,\boldsymbol{v})}
\eeq
is entire.

One can see that the above definition implies that 
$\mathcal{Q}_{\{ \eta\},\{ \xi\}}(u|\,\boldsymbol{v})$
vanishes at $u=v_j$ and hence the roots $v_j$ always satisfy the weak GBE \eqref{bethe}.
However, the definition of the strong GBE is slightly more restrictive as it provides additional equations in the case of multiple roots (\textit{ie} $v_j=v_k$ for some $j\not= k$).

%%%%%%%%%%%%%%%%%%%%%%%%%%%%%%%%%%%%%%%%%%%%%%%%%%%%%%%%%%%%%%%%%%%%%%%%%%%%%%%%%%
\subsection{General properties of solutions}
In this subsection, we discuss general properties of the solutions to the 
weak and strong GBE with $F$ and $S$ given by \eqref{defFS}.
We first show that any solution to the weak, and hence to the strong GBE as well,
has necessarily real roots $v_j$.
Then, we show that for the strong GBE there are no ``double" roots (\textit{ie} $v_j\not=v_k$ for $j\not=k$).
For the both properties the repulsion condition \eqref{repulsion_cond} is crucial.
The proofs follow closely the standard ones
\cite{BogoliubiovIzerginKorepinBookCorrFctAndABA,FaddeevABALesHouchesLectures}
 for the case $K=1$.
Note that, until section \ref{Section denombrement solutions GBAE}, we ignore
the $\pi$-periodicity in the trigonometric case and assume that $v_j\in\Rbbd$. If in a pair of solutions
$\boldsymbol{v}$ and $\boldsymbol{v'}$ we have $v_n^\prime=v_n+\pi$ the solutions are treated as different ones.

\begin{prop}
\label{proposition realite des racones de Bethe}

Let $F$ and $S$ satisfy the repulsion conditions \eqref{repulsion_cond}. Then all of the coordinates $v_n$ of a solution
 $\boldsymbol{ v }=(v_1,\dots,v_M)$ to the weak GBE \Ref{bethe} are real.
\end{prop}

\noindent{\bf Proof.}
The proof goes as in \cite{BogoliubiovIzerginKorepinBookCorrFctAndABA}.
One orders the roots $v_n$ with respect to their increasing imaginary part:
\beq
    \Im v_1\leq \Im v_2\leq \ldots \leq \Im v_M.
\label{order_u}
\eeq

Then  $\Im(v_M-v_m)\geq 0\ \forall m$  and therefore $\abs{S(v_M-v_m)}\geq1$
due to the repulsion condition \Ref{repulsion_cond}. The GBE \Ref{bethe}
then imply $\abs{F(v_M)}\leq1$ that is $\Im v_M \leq 0$, as $F$ satisfies the repulsion condition.
Similarly, setting $n=1$, we obtain that $\Im(v_1-v_m)\leq 0\ \forall m$
leading to $\abs{S(v_1-v_m)}\leq1$ and therefore $\abs{F(v_1)}\geq1$,
that is to say $\Im v_1\geq0$.

The two inequalities $\Im v_M \leq 0$ and $\Im v_1\geq0$ along with \eqref{order_u} lead to $\Im v_n=0$ for all $n$. \endproof

\begin{prop} {\rm [The exclusion principle]}\label{Proposition repoussement des racines}
Let $F, S$ be as in \eqref{defFS},
and  $\boldsymbol{ v }$ be a solution to the strong GBE. Then
the coordinates of $\boldsymbol{ v }$ are pairwise distinct, \textit{ie} $v_j\not=v_k$ for $ j \not= k$.
\end{prop}

\noindent{\bf Proof.} Again, we follow closely a proof from \cite{BogoliubiovIzerginKorepinBookCorrFctAndABA} for the case $K=1$.
Assume that there exists a solution $\boldsymbol{ v }=(v_1,\dots,v_M)$ to the strong GBE with a double root $v_k=v_j$  for some $j\neq k$. 
The condition for the expression \Ref{QoverQ} to be entire 
implies that $u=v_k$ is a double root of
$\mathcal{Q}_{\{ \eta\},\{ \xi\}}(u|\,\boldsymbol{ v } )$ \eqref{definition Q} and thus
\beq\label{eq:bethev2}
   \left.\frac{\rd  }{\rd u } \cdot  \mathcal{Q}_{\{ \eta\},\{ \xi\}}(u|\,\boldsymbol{ v } )\right|_{u=v_k}     =0.
\eeq

By  proposition \ref{proposition realite des racones de Bethe}, all the roots $v_k$ are real. 
Therefore,
\be
   \alpha(v_k\mid \{\xi \}) \prod_{\ell=1}^{K} \prod_{m=1}^{M} \psi(v_k-v_m+\eta_\ell) \not=0,
\ee   
and one can apply  the weak GBE \eqref{bethe} to recast the derivative of $\mathcal{Q}_{\{ \eta\},\{ \xi\}}$ as
\begin{multline}
   \left.\frac{\rd   }{\rd u}\cdot \mathcal{Q}_{\{ \eta\},\{ \xi\}}(u|\,\boldsymbol{ v } ) \right|_{u=v_k}
= - \alpha(v_k\mid \{\xi \}) \prod_{\ell=1}^{K} \prod_{m=1}^{M} \psi(v_k-v_m+\eta_\ell) \\
 \times  \left\{
\sum_{n=1}^{N} \frac{ \psi(2 i \Im \xi_n ) }{ |\psi(v_k + \xi_n )|^2  } +
\sum_{\ell=1}^{K} \sum_{m=1}^{M} \frac{ \psi(2i\Im \eta_{\ell} ) }{ |\psi(v_k-v_m + \eta_{\ell} )|^2  }
\right\} \; .
\end{multline}

The factor appearing in the second line is purely imaginary with a strictly positive imaginary part and the pre-factor does not vanish. This contradicts the assumption that $\mathcal{Q}^{\prime}_{\{ \eta\},\{ \xi\}}(v_k\mid \boldsymbol{v})=0$.
\endproof
\vskip2mm

We have thus proven that in the repulsive case the strong GBE are equivalent to the weak GBE \eqref{bethe}
supplemented with the condition $v_j\not= v_k$ for $j\not=k$. From now on, we can focus on the analysis of the weak GBE \eqref{bethe} only, simply discarding the solutions with multiple roots.

From the physical point of view, the solutions $(v_1,\ldots,v_M)$ differing by a permutation of roots $v_j$
are indistinguishable, defining the same physical state. In the repulsive case, when all roots are real and distinct,
it is sufficent then to consider only the {\em ordered} solutions that satisfy
\beq\label{ordered-sln-v}
      v_1<v_2<\ldots<v_M.
\eeq

In this text, however, we do not assume condition \Ref{ordered-sln-v}, unless stated otherwise.
The solutions satisfying \Ref{ordered-sln-v} are always explicitely referred to as {\em ordered} solutions.

%%%%%%%%%%%%%%%%%%%%%%%%%%%%%%%%%%%%%%%%%%%%%%%%%%%%%%%%%%%%%%%%%%%%%%%%%%%%%%%%%%%%%%%%%%
\subsection{Logarithmic Bethe equations}

The logarithmic Bethe equations \cite{BogoliubiovIzerginKorepinBookCorrFctAndABA,FaddeevABALesHouchesLectures}
 provide a way of rewriting the Bethe equations in such a way that a given solution $\boldsymbol{v} =(v_1,\dots,v_M)$ becomes labelled by a set of $M$ integers $I_1,\dots,I_M$.
These equations (logBE) are obtained by taking the logarithm of \eqref{bethe}:
\beq\label{logbethe}
    \varphi(v_n)+\sum_{\substack{m=1\\ m\neq n}}^M \theta(v_n-v_m)=2\pi I_n,
    \qquad n=1\ldots M,
\eeq
with $\varphi$ and $\theta$ related to the logarithms of $F$ and $S$
\beq
    \varphi(u)\equiv\ri\ln F(u)   \quad \text{and} \quad \theta(u)\equiv\ri\ln S(u).
\label{defintion phi et theta}
\eeq

We fix the branch of the logarithm of $S(u)$ by demanding that $\theta(0)=\pi K$,
which is possible due to the symmetry \eqref{symmetry condition on etas}.
As for the branch of $\ln F(u)$,  we just assume that it is fixed in some arbitrary way.
We also stress that
$\varphi$ and $\theta$ are strictly increasing functions

\beq
\varphi^{\prime}(u)=-i\sum_{n=1}^{N} \frac{ \psi(2 i \Im \xi_n ) }{ |\psi(u + \xi_n )|^2  } >0  
\quad \text{and} \quad
\theta^{\prime}(u)=-i\sum_{\ell=1}^{K}  \frac{ \psi(2i\Im \eta_{\ell} ) }{ |\psi(u + \eta_{\ell} )|^2  }>0.
\label{propriete positivite phi prime et theta prime}
\eeq

Moreover, due to \eqref{symmetry condition on etas}, $\theta^{\prime}$ is even.

To be able to use the above correspondence of the solutions to GBE and $M$-tuples of integers
we need to describe the class of admissible integers $I_n$ that can arise in the right-hand side of
\Ref{logbethe} and to establish a bijection between those $M$-tuples and
the solutions to GBE.

The construction is different in the trigonometric and rational cases and is described in the next section. In the rest of the current section, we present the properties of the integers $I_n$ that are
shared by the trigonometric and rational cases. 

\begin{prop} {\rm [Monotonicity principle]}.
\label{proposition monotonicity principle}
Under the repulsion conditions \eqref{repulsion_cond},
a growing sequence of roots $v_n$
of the logBE \Ref{logbethe} corresponds to
a growing sequence of integers $I_n$, and vice versa:
\beq
    v_k>v_j \quad \Longleftrightarrow\quad I_k>I_j.
\eeq
\end{prop}

\noindent{\bf Proof.} 
Suppose that $v_2>v_1$. From the strict monotonicty \eqref{propriete positivite phi prime et theta prime} of $\varphi(u)$, resp.\ $\theta(u)$,
it follows that $\varphi(v_2)>\varphi(v_1)$, resp.\
$\theta(v_2-v_1)>\theta(v_1-v_2)$ and $\theta(v_2-v_m)>\theta(v_1-v_m)$
for $m>2$. Adding all the inequalities together we obtain $I_2>I_1$.

Now assume that $I_2>I_1$. If $v_1=v_2$ then clearly $I_1=I_2$, that contradicts the assumption.
Hence $v_1\not=v_2$. Also, the variant $v_2<v_1$ is impossible as, due to the above paragraph, it would imply that $I_1>I_2$. Therefore $v_2>v_1$.
\endproof

\begin{prop}
\label{proposition unicite des solutions}

Let $\boldsymbol{I}\equiv(I_1,\dots, I_M)$ be $M$ integers such that there exists a solution ${\boldsymbol{v}}=(v_1,\dots,v_M)$ to the logBE \eqref{logbethe}. Then ${\boldsymbol{v}}$ is the only solution corresponding to the same $\boldsymbol{I}$.

\end{prop}

\noindent{\bf Proof.}
Note that the logBE \eqref{logbethe} appear as the extremum condition
$w_n\equiv\dd W/\dd v_n=0$ for the potential $W(\boldsymbol{v})$ on $\mathbb{R}^M$:
\beq
    W(\boldsymbol{v})\equiv\sum_{n=1}^M \Phi(v_n)
    +\frac12 \sum_{m,n=1}^M \Theta(v_n-v_m)
    -2\pi\sum_{n=1}^M I_n v_n
    +\pi(M-1)  K \sum_{n=1}^M v_n,
\label{definition potential W}
\eeq
where $\Phi$ and $\Theta$ are the integrals of $\varphi$ and $\theta$ \eqref{defintion phi et theta}
\beq
    \Phi(u)\equiv \int_0^u \varphi(t)\,\rd t,
 \quad \Theta(u)\equiv \int_0^u\theta(t)\,\rd t \; .
\label{big theta and phi}
\eeq

The calculation of $w_n$ uses the identity $\theta(u)+\theta(-u)=2\theta(0)=2\pi K$
that follows from $\theta^{\prime}$ being an even function.

To prove the uniqueness of solutions to \eqref{logbethe} for a given choice of $\boldsymbol{I}$, it suffices to show that the potential is strictly convex, and therefore, if it admits a minimum, the latter is unique.
To prove the strict convexity of $W$ it is enough to check that the Hessian of $W$
\begin{align}
    \frac{\dd^2 W}{\dd v_n\dd v_p}
    =\frac{\dd w_n}{\dd v_p}
    &=\delta_{np}\left(\varphi'(v_n)
    +\sum_{m=1}^M \theta'(v_n-v_m)\right)
    -\theta'(v_n-v_p).
\label{hessian}
\end{align}
is a positively definite matrix. Evaluating the quadratic form of the Hessian \Ref{hessian}
on an arbitrary nonzero vector $\boldsymbol{g}\in \Rbbd^M$ we obtain
\begin{align}
  \sum_{n,p=1}^M \frac{\dd^2 W}{\dd v_n\dd v_p}\,g_ng_p
  &=\sum_{n=1}^M g_n^2\left(\varphi'(v_n)
    +\sum_{m=1}^M \theta'(v_n-v_m)\right)
    -\sum_{n,p=1}^M\theta'(v_n-v_p)\,g_ng_p    \notag\\
  &=\sum_{n=1}^M \varphi'(v_n)g_n^2
   +\sum_{n>m}^{M} \theta'(v_n-v_m)(g_n-g_m)^2
   \geq \sum_{n=1}^M \varphi'(v_n)g_n^2 >0.
\end{align}

Here we used that $\varphi^{\prime}>0$ and $\theta^{\prime}>0$ on $\Rbbd$,
and that $\theta^{\prime}$ is even.
\endproof

In the $K=1$ case the potential $W$ is of course the well-known Yang-Yang function 
\cite{BogoliubiovIzerginKorepinBookCorrFctAndABA}.

%%%%%%%%%%%%%%%%%%%%%%%%%%%%%%%%%%%%%%%%%%%%%%%%%%%%%%%%%%%%%%%%%%%%%%%%%%%%%%%%
\section{Solvability of the generalized Bethe equations}
\label{Section criteres de solvabilite}

In this section we derive the necessary and sufficient conditions on the integers $I_j$
for the corresponding GBE to have a (unique, by proposition \ref{proposition unicite des solutions})
solution. The problem of counting of the solutions to GBE will be thus reduced to the enumeration
of the allowed sets of integers. The description of those sets is different in the trigonometric and rational cases.

\subsection{Trigonometric case}

It is useful to introduce an auxiliary function:
\beq
\chi(u,c) = i \ln \frac{ \sin(u+ic) }{ \sin(u-ic) }= \int\limits_{-\frac{\pi}{2} }^{ u } \frac{ \sinh (2c) }{ \sin(v+ic)\sin(v-ic) }\, \text{d} v\,,
\label{definition fonction chi}
\eeq
fixing the branch of logarithm by the condition $\chi(-\pi/2)=0$.
Note that $u \mapsto \chi(u,c)$ is $\pi$ quasi-periodic:
\beq\label{eq:chi-quasip}
    \chi(u+\pi k, c)=2\pi k+\chi(u,c), \quad k\in\Zbbd\; \;  \; \text{and} \; \text{such} \; \text{that}\quad
\chi(-u,c)+\chi(u,c)=2\pi \;.
\eeq
 Moreover, $\chi(u,c)$ grows monotonously on the real axis and its integral
has the asymptotic behavior
\beq
 \int_0^u \chi(t,c)\,\rd t  =u^2+O(u),\qquad u\rightarrow\pm\infty.
\label{comportement asymp primitive chi}
\eeq

The functions $\chi(u,c)$ allow us to provide a more explicit representation for $\varphi$ and $\theta$:
\begin{subequations}\label{eq: def phi-theta trigo}
\begin{align}\label{definition phi trigo}
    \varphi(u)&= 2\pi\varphi_\infty
    +\sum_{n=1}^N \chi\left(u+\Re(\xi_n),\Im(\xi_n) \right), \quad \text{where}\quad F_\infty=\re^{-2\pi\ri\varphi_\infty},\\
\label{definition theta trigo}
    \theta(u)&=
    \sum_{j=1}^K \chi(u+\Re(\eta_j),\Im(\eta_j)).
\end{align}
\end{subequations}

One can check using \eqref{eq:chi-quasip} and \eqref{symmetry condition on etas} that indeed one has $\theta(0)=\pi K$.

We are now in position to prove the existence of solutions in the trigonometric case.

\vspace{5mm}

\begin{prop}\label{proposition existence solutions trigonometrique}
Let $\boldsymbol{I}=(I_1,\ldots,I_M)\in\Zbbd^M$ be any sequence of $M$ integers. Then, in the trigonometric case where $\psi(u)=\sin(u)$, there exists a unique solution to the logBE.
\end{prop}

\noindent{\bf Proof.} 
We have already seen in proposition \ref{proposition unicite des solutions} that, if a solution exists, then it is unique as it provides the minimum for a strictly convex potential $W$ \eqref{definition potential W}. To prove the existence of solutions, it is thus enough to show that $W$ grows at infinity.
Indeed, $W$ is then bounded from below and hence has a local minimum.

It follows from \eqref{comportement asymp primitive chi} that $\Phi$ and $\Theta$, the integrals of $\varphi$ and $\theta$, have the asymptotic behavior
\beq
    \Phi(u)= \int_0^u \varphi(t)\,\rd t=Nu^2+O(u)\; , \qquad \Theta(u)= \int_0^u\theta(t)\,\rd t
    =Ku^2+O(u) \quad \text{for} \quad
    u\rightarrow\pm\infty \;.
\label{asymptotics big theta and phi}
\eeq

Hence, it follows that  $W( \boldsymbol{v} )$ has the asymptotics
\beq
   W(\boldsymbol{v})=
   N\sum_{n=1}^M v_n^2 + \frac{K}{2} \sum_{m,n=1}^M (v_n-v_m)^2
   +O(\max_{j} |v_j|).
\eeq

The potential $W$ thus grows in every direction at infinity, and hence $W$ has a local minimum.
The existence of solutions to \eqref{logbethe} is thus proven. \endproof
\vskip2mm

So far we ignored completely the periodicity of trigonometric functions. However,
the periodicity $F(u+\pi)=F(u)$, $S(u+\pi)=S(u)$ means that two different solutions
$\boldsymbol{v},\boldsymbol{v'}\in\Rbbd^M$ of the logarithmic Bethe equations \eqref{logbethe}
are equivalent as solutions of the weak GBE \eqref{bethe} if
$\boldsymbol{v}-\boldsymbol{v'}\in(\pi\Zbbd)^M$.
To obtain a characterization of the set of distinct solutions 
one needs to express the equivalence of the vectors $\boldsymbol{v}$ modulo the lattice $(\pi\Zbbd)^M$
in terms of the integers $\boldsymbol{I}$. The answer is given below.

Denote 
\beq\label{eq:def-ab}
    a\equiv N+(M-1)K,\qquad b\equiv-K.
\eeq

\begin{prop}\label{prop: periods of I}
Let two solutions $\boldsymbol{v},\boldsymbol{v'}\in\Rbbd^M$ of the logBE \eqref{logbethe} differ by the shift of a single root $v_n$ by $\pi$:
\beq\label{eq:shift-v}
       v_m^\prime=v_m+\pi\delta_{mn},\qquad m=1,\ldots,M.
\eeq

Then for the corresponding integer vectors we have
\beq\label{eq: shift I}
      I_m^\prime=I_m+a\d_{mn}+b(1-\d_{mn}),\qquad m=1,\ldots,M.
\eeq

\end{prop}

\noindent{\bf Proof.} The statement follows from the logBE \eqref{logbethe}
and from the quasi-periodicity conditions
\beq\label{eq:phi-quasip}
    \varphi(u+\pi)=2\pi N+\varphi(u),\qquad      \theta(u+\pi)=2\pi K+\theta(u).
\eeq
resulting from \Ref{eq:chi-quasip} and \Ref{eq: def phi-theta trigo}.
\endproof
\vskip2mm

\begin{cor}
The set of distinct {\em ordered} in the sense \Ref{ordered-sln-v} solutions of the trigonometric weak GBE
\Ref{bethe} is in one-to-one correspondence with the equivalence classes $\Ical\in\Zbbd^M/\gbf$ 
of integer vectors $\boldsymbol{I}\in\Zbbd^M$
factorised over the lattice of the periods $\gbf$ generated by the basis $\gamma_n$ $(n=1,\ldots,M)$:
\beq\label{eq: def basis gamma}
   (\g_n)_m=a\d_{mn}+b(1-\d_{mn}),\qquad \gbf = \left\{\sum_{n=1}^M c_n\g_n,\quad c_n\in\Zbbd\right\}.
\eeq

\beq
   \Ibf \approx \boldsymbol{I'}\quad\Longleftrightarrow\quad \Ibf-\boldsymbol{I'}\in\gbf.
\eeq
\end{cor}

Having thus characterized the distinct ordered solutions to the {\em weak} GBE in terms of the integer vectors $\Ibf$ we proceed to characterize the solutions to the {\em strong} GBE.

\begin{defin}\label{def: regular class I} 
An equivalence class $\Ical\in\Zbbd^M/\gbf$ is called {\em regular} if all its representatives $\Ibf$ lie off diagonals of $\Zbbd^M$, that is $I_m\neq I_n$ for $m\neq n$.
\end{defin}

\begin{theo}\label{theorem charactersation solution cas trig}
The set of distinct ordered solutions to the strong GBE in the trigonometric case is in a one-to-one correspondence with the set of regular equivalence classes $\Ical$.
\end{theo}

\noindent{\bf Proof.} The exclusion principle (proposition \ref{Proposition repoussement des racines})
demands to disregard the solutions $\boldsymbol{v}$ of the logBE \Ref{logbethe}
having multiple roots. More precisely, we discard the whole equivalence class $\Vcal\in\Rbbd^M/(\pi\Zbbd)^M$ if 
it contains a representative $\vbf\in\Vcal$ such that $v_m=v_n$ for some $m\neq n$.
Indeed, in such a case, by the monotonicity principle (proposition \ref{proposition monotonicity principle}),
for the corresponding integer vector $\Ibf$ we have $I_m=I_n$ for the same $m\neq n$.
Eliminating the equivalence classes generated by such representatives $\Ibf$ we are left with
the regular classes $\Ical$. Vice versa, by the same monotonicity principle, if $\Ical$ is a regular class, 
all its representatives $\Ibf$ give rise to solutions $\vbf$ with distinct coordinates. 
\endproof

%%%%%%%%%%%%%%%%%%%%%%%%%%%%%%%%%%%%%%%%%%%%%%%%%%%%%%%%%%%%%%%%%%%%%%%%%%%%%%%%%%%%%%%%%%%%%%%%%%%%
\subsection{The rational case}

We now discuss the existence of solutions in the rational case. We will show that the condition of
existence of solutions leads to a quite different description of the set of admissible integers
$\boldsymbol{I}$ leading to distinct solutions of the rational GBE.

We start by introducing the rational analog of the function $\chi(u,c)$:
\beq
\rho(u,c) = i\ln \left( \frac{ u+ic}{u-ic} \right) = \int\limits_{-\infty}^{u} \frac{ 2c}{ v^2+c^2} \text{d} v \; , \quad c >0\;.
\eeq

It satisfies $\rho(u,c)+\rho(-u,c)=2\pi$ and has asymptotics
\beq
    \rho(u,c) = \left\{\begin{array}{rl}
          0-2cu^{-1}+O(u^{-3}),&\quad u\rightarrow-\infty\\
          2\pi-2cu^{-1}+O(u^{-3}),&\quad u\rightarrow+\infty
    \end{array}\right. \; .
\eeq

The integral of $\rho$ is explicitly computable and reads
\beq\label{asymptotique integrale rho}
    \int_0^u \rho(t,c)\,\rd t
            =\pi u+2u\arctan\frac{u}{c}-c\ln\left(1+\frac{u^2}{c^2}\right)
        =\left\{\begin{array}{rl}
          -2c\ln\abs{u}+O(1),&\quad u\rightarrow-\infty \vspace{2mm}\\
          2\pi u-2c\ln u+O(1),&\quad u\rightarrow+\infty
    \end{array}\right.  \; .
\eeq

In the rational case, $\varphi$ and $\theta$ are given by
\beq
    \varphi(u)= 2\pi\varphi_\infty
    +\sum_{n=1}^N \rho\left(u+\Re(\xi_n),\Im(\xi_n) \right) \;,
\qquad
    \theta(u)=
    \sum_{j=1}^K \rho(u+\Re(\eta_j),\Im(\eta_j))\; .
\eeq

Again $\theta(0)=\pi K$ follows from \eqref{symmetry condition on etas}. We also remind that $F_{\infty}=\text{e}^{-2i\pi \varphi_{\infty}}$.

\vspace{5mm}

\begin{prop}
\label{theorem characterisation solutions cas rationel}
Assume that the phase $\varphi_{\infty}$ is such that
\beq
    -1 < (M+1)\varphi_\infty <0.
\label{eq:interval-phi}
\eeq
then there exists a solution to the logarithmic GBE in  the rational case if and only if
for any subset $\Jcal\subset\{1,\ldots,M\}$ of cardinality $\abs{\Jcal}=r$ with $r=1,\dots,M$,
the $M$ integers $(I_1,\dots,I_M)$ satisfy the set of $2^M$ inequalities:
\beq\label{eq:ineqI}
    \frac{r(r-1)}{2}\,K
    \,\,\leq\,\, \sum_{j\in\Jcal} I_j \,\,\leq\,\,
    rMK-\frac{r(r+1)}{2}\,K+rN-1.
\eeq
\end{prop}

\begin{cor} {\em(follows from the above result combined with propositions \ref{proposition realite des racones de Bethe}-\ref{proposition unicite des solutions}).}
The distinct ordered solutions to the rational generalized Bethe equations are in one-to-one correspondence with
monotonously increasing sequences of $M$ integers $I_1<\dots < I_M$ satisfying the set of $2^M$ constraints \eqref{eq:ineqI}.
\end{cor}

\noindent{\bf Proof} [\textit{Proposition} \ref{theorem characterisation solutions cas rationel}].
Similarly to  the proof of proposition \ref{proposition existence solutions trigonometrique}, it is enough
to prove the divergence of the potential $W$ at infinity so as to ensure the existence of solutions. Reciprocally,
if one is able to show that the potential is unbounded from below, then it cannot have a local minimum. Indeed due to the strict convexity of $W$, the local minimum would be global, contradicting the unboundedness of $W$.

We first assume that the set of $2^M$ conditions \eqref{eq:ineqI} holds
and prove that $W$  goes to $+\infty$ along any ray. This is enough to prove that $W$ admits at least one minimum.

Next, we assume that at least one of the $2^M$ conditions \eqref{eq:ineqI} does not hold. Then we construct a ray along which the potential goes to $-\infty$. As discussed above, this implies that $W$ cannot have any local minimum.

\vspace{1mm}
Suppose that the M integers $I_1,\dots,I_M$ satisfy to the constraints \eqref{eq:ineqI}.
It follow from the asymptotics \eqref{asymptotique integrale rho} of the antiderivative of $\rho(u,c)$, that $\Phi(u)$ and $\Theta$ have the asymptotics
\be
    \Phi(u)\equiv \int_0^u \varphi(t)\,\rd t=
       \left\{\begin{array}{rcl}
          2\pi\varphi_\infty u-2\left(\sum_{j=1}^N \Im(\xi_j) \right)\ln\abs{u}+O(1), &\quad& u\rightarrow -\infty \vspace{2mm}\\
          2\pi(\varphi_\infty+N)u-2\left(\sum_{j=1}^N \Im(\xi_j) \right)\ln u+O(1), &\quad& u\rightarrow +\infty
       \end{array}\right. \;,
%\label{eq:asPhi}
%
\ee
\be
    \Theta(u)\equiv \int_0^u \theta(t)\,\rd t=
       \left\{\begin{array}{rcl}
          -2\left(\sum_{j=1}^N \Im(\eta_j) \right)\ln\abs{u}+O(1), &\quad& u\rightarrow -\infty \vspace{2mm} \\
         2\pi K u-2\left(\sum_{j=1}^N \Im(\eta_j) \right)\ln u+O(1), &\quad& u\rightarrow +\infty
       \end{array}\right. \;.
%
%\label{eq:asTh}
\ee

Up to a permutation of the $I_j$'s, $W$ is a symmetric function of the $v_k$'s. Therefore we
permute the coordinates of ${\boldsymbol{v}}$ with $\sigma \in \mathfrak{S}_M$ in such a way that ${\boldsymbol{v}}_{\sigma}\equiv(v_{\sigma(1)},\dots,v_{\sigma(M)}) = t \boldsymbol{u}$, where $t>0$ and
${\boldsymbol{u}} \; : \; \sum_{i=1}^{M} u_i^2=1$ has its coordinates ordered in the following way
\beq
\boldsymbol{u}= ( \underbrace{u_1, \dots, u_1}_{\la_1}\, , \dots\, ,\,  \underbrace{u_s,\dots, u_s}_{\la_s} )
\quad \text{with} \quad u_1<u_2<\dots<u_q\leq0<u_{q+1}<\dots <u_s \; .
\label{definition ordering u coords}
\eeq
Note that we have $\la_p>0\,$ for $p=1 \dots s$ and $M=\la_1+\dots+\la_s$. The above permutation  of the coordinates of $\boldsymbol{v}$ results in a permutation of the integers $I_j \mapsto I_{\sigma(j)}$.

For further convenience we introduce the shorthand notations
\beq
w_k =  \sum_{p=1}^{k} \lambda_p \quad \text{and} \quad
J_{\ell} = \sum\limits_{j=w_{\ell-1} +1 }^{ w_{\ell} } I_{\sigma(j)} \;,
\label{definition nombre omega k}
\eeq
and agree upon $w_0=0$. After some algebra, we recast $W$ in the form
\beq
W(\boldsymbol{v})=\sum_{k=1}^{s} \la_k \Phi(t u_k) + \frac{1}{2}\sum_{i,j=1}^{s} \lambda_i\lambda_j \Theta(tu_i-tu_j)
+ \pi (M-1)K \sum_{k=1}^{s} \lambda_k t u_k - 2\pi \sum_{k=1}^{s} t u_k J_k \;.
\eeq
Here, we can already send $t \rightarrow +\infty$ and compute the asymptotic behavior of all functions
without problem:

\beq
\Phi(t u_k) =\left\{ \begin{array}{cc}
                    2\pi \varphi_{\infty} tu_k + \text{O}( \ln t)\; , & k \in \{1,\dots, q\} \vspace{2mm},\\
                 2\pi (\varphi_{\infty}+N) tu_k  + \text{O}( \ln t) \; , & k \in \{q+1,\dots, s\},

\end{array} \right.
\eeq

\beq
\sum_{i,j=1}^{s} \lambda_i\lambda_j \Theta(tu_i-tu_j)
= 2\pi K t  \sum_{1\leq i<j\leq s}^{} \lambda_i\lambda_j (u_j-u_i) +\text{O}(\ln t) \; .
\eeq

As a result, the asymptotics can be recast in the form
\beq
W(t\boldsymbol{u})= 2\pi t
\left( \sum_{k=1}^{q} u_k \tau^{(-)}_k  +  \sum_{k=q+1}^{s} u_k \tau_k^{(+)} \right)  +  \text{O}(\ln t) \; ,
\label{somme asympt W rat}
\eeq
where the two sequences $\tau^{(\pm)}_k$ read
\begin{align*}
\tau^{(-)}_k &=\lambda_k \left[  \varphi_{\infty} + \frac{K}{2} (\lambda_k-1) + Kw_{k-1} \right]
 -J_k \; , \\
\tau^{(+)}_k &=\lambda_k
\left[ \varphi_{\infty}+ N +M K  - \frac{K}{2} (\lambda_k+1) - K  (M-w_k)   \right]
 -J_k \; .
\end{align*}

The claim will follow as soon as we prove that the sum in the brackets in \eqref{somme asympt W rat} is strictly
positive. We first focus on the case where $u_q\not=0$. Then the ``$(+)$" and  ``$(-)$" partial sums are both strictly positive. Indeed we observe that $\sum_{k=1}^{q} u_k \tau_k^{(-)}$ can be re-cast as:
\beq
\sum_{k=1}^{q} u_k \tau_k^{(-)}=   \sum_{k=1}^{q-1} (u_k-u_{k+1}) \left(\sum_{j=1}^{k}\tau_j^{(-)} \right)
 + u_q \sum_{j=1}^{q} \tau_j^{(-)}  \; .
\eeq
There, we have from the very definition of the ordering of the $u_k$'s that $u_k-u_{k+1}<0$ and $u_q < 0$.
Also
\begin{align*}
\sum_{j=1}^{k}\tau^{(-)}_j &= \varphi_{\infty}\sum_{p=1}^{k} \lambda_p  +
\frac{K}{2}\sum_{p=1}^{k} \lambda_p (\lambda_p-1) + K\sum_{j<p}^{k}\lambda_p \lambda_j
 -\sum_{j=1}^{k} J_{j}  \notag\\
&=\varphi_{\infty} w_k
-\frac{K}{2} w_k + \frac{K}{2}\sum_{j,p=1}^{k}\lambda_p \lambda_j
 -\sum_{j=1}^{w_k} I_{\sigma(j)} \\
& \leq  \varphi_{\infty}w_k + \frac{K}{2}w_k(w_k-1) - \frac{K}{2} w_k (w_k-1)= \varphi_{\infty}w_k <0
\end{align*}
where we have used the $lhs$ of \eqref{eq:ineqI} and the definition of $w_k$ \eqref{definition nombre omega k} so as to obtain the last line. It thus follows that $\sum_{k=1}^{q} u_k \tau^{(-)}_k>0$.

It now remains to show that $\sum_{k=q+1}^{s} u_k \tau_k^{(+)}>0$. Very similarly, we decompose
\beq
\sum_{k=q+1}^{s} u_k \tau_k^{(+)}=   \sum_{k=q+2}^{s} (u_k-u_{k-1}) \left(\sum_{j=k}^{s}\tau_j^{(+)} \right)
 + u_{q+1} \cdot  \sum_{j=q+1}^{s} \tau_j^{(+)}  \; .
\eeq
As $u_k-u_{k-1}>0$ and $u_{q+1} >0$, we ought to check the sign of the partial sums involving the $\tau_k^{(+)}$.
We set $\widetilde{w}_k=\sum_{j=k}^{s} \lambda_k=M-w_{k-1}$ and get
\beq
\sum_{j=k}^{s}\tau_j^{(+)}= \widetilde{w}_k (\varphi_{\infty}+ N +M K ) - \frac{K}{2} \widetilde{w}_k(\widetilde{w}_k+1) - \sum_{j=\widetilde{w}_{k-1}+1}^{M} I_{\sigma(j)}
\geq \widetilde{w}_k \varphi_{\infty} +1 >0 \; .
\eeq
The last inequality follows by applying the $rhs$ part of \eqref{eq:ineqI}.

Now we assume that $u_q=0$. As $\boldsymbol{u}$ lies on the unit sphere it has to have non-zero coordinates. This means that necessarily $s>1$. In its turn this ensures that even if the contribution of $u_q$ to the asymptotics is zero, there are still other $u_{\ell}$'s such that taken as a whole, the sum in the brackets in \eqref{somme asympt W rat} is strictly positive.

Therefore, provided that the M  integers $I_j$, $j=1,\dots,M$ satisfy the $2^M$ inequalities \eqref{eq:ineqI},
we see that the potential $W(\boldsymbol{v})$ grows in all directions. This implies that it admits at least one local minimum, and hence that there exists a solution to the logarithmic Bethe equation \eqref{logbethe}
associated with these integers.

Reciprocally, suppose that one of the $2^M$ inequalities \eqref{eq:ineqI} is not satisfied.
We distinguish between two cases. First suppose that there exists a subset $\mathcal{J}$
of cardinality $r\in\{1,\dots,M\}$ such that the $lhs$ part of inequality \eqref{eq:ineqI}
is not satisfied. Then, as the inequalities only involve integers, we necessarily have
\beq
 K \frac{r(r-1)}{2} -1\geq \sum_{k=1}^{r} I_{\sigma(k)} \;.
\eeq
Here $\sigma \in \mathfrak{S}_M$ is a permutation that maps $\{ 1,\dots, r\}$  onto $\mathcal{J}$. We choose to send the vector $\boldsymbol{v}$ to infinity along the ray
\beq
v_{\sigma(k)} = -\frac{t}{\sqrt{r}} \; , \; \text{for}\;  k =1,\dots, r \quad \text{and} \quad
v_{\sigma(k)} = 0 \; ,  \; \text{for}\; k =r+1,\dots, M \; .
\eeq
Then according to \eqref{somme asympt W rat}, we get
\beq
W(\boldsymbol{v}) = -\frac{2\pi t}{\sqrt{r}}  \tau^{(-)}_1 + \text{O}(\ln t)
\eeq
where
\beq
\tau^{(-)}_1 = r \varphi_{\infty} + \frac{K}{2} r (r-1)   -\sum_{k=1}^{r} I_{\s(k)}  \geq r \varphi_{\infty} +1  >0 \;.
\eeq
This shows that $W$ is unbounded from below and hence cannot have any minimum. Similarly, if there exists a subset $\mathcal{J}$ of cardinality $r\in\{1,\dots,M\}$  such that the $rhs$ of inequality \eqref{eq:ineqI}
is not satisfied, then
\beq
\sum_{k=1}^{r} I_{\sigma(k)} \geq r M K - \frac{r(r+1)}{2}K + rN \;.
\eeq
Here $\sigma \in \mathfrak{S}_M$ is any permutation that maps $\{ 1,\dots, r\}$  onto $\mathcal{J}$. We choose to
send the vector $\boldsymbol{v}$ to infinity along the ray
\beq
v_{\sigma(k)} = \frac{t}{\sqrt{r}} \; , \; \text{for} \;\; k =1,\dots, r \quad \text{and} \quad
v_{\sigma(k)} = 0 \; , \; \text{for} \;\; k =r+1,\dots, M \; .
\eeq
Then according to \eqref{somme asympt W rat}, we get
\beq
W(\boldsymbol{v}) = \frac{2\pi t}{\sqrt{r}}  \tau^{(+)}_2 + \text{O}(\ln t) \;.
\eeq
Paying attention to the conventions introduced in \eqref{definition ordering u coords}, we see that
$\la_1=M-r$, $\la_2=r$ and thus $w_2=M$,
\beq
\tau^{(+)}_2 = r \varphi_{\infty} + r(N + MK)-  \frac{K}{2} r(r+1)  - \sum_{k=1}^{r} I_{\sigma(k)}
\leq r \varphi_{\infty}   <0 \;.
\eeq
This shows that $W$ goes to $-\infty$ along that ray and the claim follows. \endproof

%%%%%%%%%%%%%%%%%%%%%%%%%%%%%%%%%%%%%%%%%%%%%%%%%%%%%%%%%%%%%%%%%%%%%%%%%%%%%%%%%%%%%%%%%%%%%%%%%%%%%%%%%%%%%%%%%%%%%%%%%%%%%%%%%%%%%%%%%%%%%%%%
%%%%%%%%%%%%%%%%%%%%%%%%%%%%%%%%%%%%%%%%%%%%%%%%%%%%%%%%%%%%%%%%%%%%%%%%%%%%%%%%%%%%%%%%%%%%%%%%%%%%%%%%%%%%%%%%%%%%%%%%%%%%%%%%%%%%%%%%%%%%%%%%

\section{Number of independent solutions in the trigonometric case}
\label{Section denombrement solutions GBAE}

\subsection{Properties of factor modules}

Factor modules over $\mathbb{Z}^M$ provide a convenient language for describing set of integers that are identified modulo shifts as those appearing in theorem \ref{theorem charactersation solution cas trig}.

Let $\matmz$ be the set of $M\times M$ matrices over the ring $\mathbb{Z}$. Given a matrix $G\in \matmz$, $G_{k\ell}=\gamma_{k\ell}$ , $\gamma_{k\ell}\in\mathbb{Z}$,
one defines  $\left<G\right>$ to be the {\em submodule} in $\mathbb{Z}^M$
generated by the rows of $G$. Then, the {\em factor-module}  $\mathbb{G}\equiv \mathbb{Z}^M/\left<G\right>$  is the set of equivalence classes of the integer vectors
$(I_1,\ldots,I_M)$ differing by a vector from $\left<G\right>$.
In other words, $\mathbb{G}$ is generated by the {\em generators} forming a column vector $\boldsymbol{e}=(e_1,\dots ,e_m)^t$
subject to the {\em relations}
\beq\label{eq:Ge=0}
    G\boldsymbol{e}=0, \quad \text{or} \; \text{explicitly} \quad     \sum_{n=1}^M \g_{mn}e_n=0,
    \qquad m=1,\ldots,M \; .
\eeq

\vspace{2mm}

The factor module $\mathbb{G}$ relevant for counting the solutions to the generalized Bethe equations,
is given by a matrix $G$ with entries
\beq\label{def-g}
   \g_{mn}=\left\{\begin{array}{rcl}
          a,&\quad& n=m,\\
          b,&\quad& n\neq m,
   \end{array}\right.
\qquad \text{or} \qquad        \g_{mn}=a\d_{mn}+b(1-\d_{mn})\;,
\eeq
provided that one takes $a=-K$ and $b=N+(M-1)K$.

Indeed, the above factor-module $\mathbb{G}$ is  identical to the set of equivalence classes
of the integer sequences $\boldsymbol{I}$ with respect to the transformations
\Ref{eq: shift I}.

\vspace{5mm}
The strategy we adopt for counting the number of solutions of the generalized  Bethe equations \eqref{bethe}
in the trigonometric case is as follows. We consider a factor module $\mathbb{G}$ defined by the matrix $G$
given in \eqref{def-g}. We start by determining the total number of elements belonging to $\mathbb{G}$.
Then, in order to relate $\# \mathbb{G}$ to the number of distinct solutions as described in theorem \ref{theorem charactersation solution cas trig} we should discard all sequences ${\bf I}=\sum_{k=1}^{M}I_k e_k \in \mathbb{G}$ which have at least one repeating component $I_j=I_k$, for some $j\not=k$. Finally, as the independent solutions of the generalized Bethe equations are in bijection with the monotonously growing sequences $I_1<\dots<I_M$, we have to divide this last result by the number of allowed permutations so as to obtain the number of solutions. \vspace{5mm}

Recall that the permutation group acts in a natural way on the module $\Zbbd^M$:
\beq
    \s: (I_1,\ldots,I_M)\mapsto (I_{\s(1)},\ldots,I_{\s(M)}).
\eeq
It is obvious that the  submodule $\left<G\right>$ given by the matrix G \eqref{def-g}
is invariant under the above action of $\mathfrak{S}_M$. The induced action of $\mathfrak{S}_M$
can be then defined on the factor-module $\mathbb{G}$.

An {\em orbit} of a point ${\bf I}=\sum_{j=1}^{M} I_{j} e_j \in\mathbb{G}$ is the set of all ${\bf I}_{\sigma}=\sum_{j=1}^{M} I_{\sigma(j)} e_j$ for $\sigma\in \mathfrak{S}_M$.
A {\em regular}, or {\em generic} orbit is an orbit that has the maximal
possible number of distinct points, that is $M!$. An example of points ${\bf I}=\sum_{k=1}^{M} I_j e_j$ not giving rise to  regular orbits are the so-called diagonal points, that is points having at least two integers coordinates equal: $I_j=I_k$ for $j\not=k$. The regular orbits of the action of $\mathfrak{S}_M$ on $\mathbb{G}$ are described by the following lemma

\begin{lemme}
The regular orbits of the factor module $\mathbb{G}$ subordinate to the period matrix $G$ \eqref{def-g} are generated by all elements ${\bf I}=\sum_{k=1}^{M}I_k e_k$ such that the integers $I_j$ form monotonous sequences  $I_1<\dots <I_M$ and the vector $(I_1,\dots,I_M)$ belongs to the fundamental domain $\mathfrak{G}$ of $\mathbb{G}$
\beq
   \mathfrak{G}=\left\{ \boldsymbol{x} \in \Rbbd^M:\ 
   x_i = t_i(a-b)+b\sum_{p=1}^{M}t_p, \quad t_i \in (0,1)   \right\}.
\eeq
\end{lemme}

\noindent{\bf Proof.}
It is clear from the previous discussion that the regular orbits can only be given by off-diagonal elements. Moreover, one can always represent $\mathbb{G}$ by points in its fundamental domain. Let ${\bf I}=\sum_{k=1}^{M}I_k e_k$  be defined in terms of a monotonous sequence of integers $I_1<I_2<\dots<I_M$ such that $(I_1,\dots,I_M)$ lies in the fundamental domain of $\mathbb{G}$.

Assume that the $\mathfrak{S}_M$-orbit of ${\bf I}$ in $\mathbb{G}=\mathbb{Z}^M/\left<G \right>$ is not regular. Then, there exist at least two distinct permutations $\sigma$ and $\sigma'$ such that
${\bf I}_{\sigma} \equiv {\bf I}_{\sigma'}$, where ${\bf I}_{\sigma} = \sum_{p=1}^{M}I_{\sigma(p)}e_p $.
In other words,
\beq
(I_{\sigma(1)},\dots,I_{\sigma(M)})=(I_{\sigma^{\prime}(1)},\dots,I_{\sigma^{\prime}(M)})+
\sum\limits_{j=1}^{M} \alpha_j (\gamma_{j1},\dots,\gamma_{jM}) \;,
\eeq
for some $\alpha_j \in \mathbb{Z}$ and $\gamma_{mn}$ given by \Ref{def-g}. 
Applying the inverse permutation, we get
\beq
(I_{1},\dots,I_{M})=(I_{\sigma^{\prime}\sigma^{-1}(1)},\dots,I_{\sigma^{\prime}\sigma^{-1}(M)})+
\sum\limits_{j=1}^{M} \alpha_{\sigma^{-1}(j)} (\gamma_{\sigma^{-1}(j)\sigma^{-1}(1)},\dots,\gamma_{\sigma^{-1}(j)\sigma^{-1}(M)}) \;.
\eeq

%\beq
%
%{\bf I}= P^{-1}_{\sigma} {\bf I}_{\sigma} = {\bf I}_{\sigma^{-1} \sigma'} + (P^{-1}_{\sigma} G P_{\sigma} ). %(P^{-1}_{\sigma}  \boldsymbol{\alpha} )
%
%\eeq
%
%
%
The period matrix $G$ \eqref{def-g} can be recast as 
\be
  G=(a-b) \text{id} + b  x\cdot y^t, 
\ee
where $x^t=y^t=(1,\dots,1)$ are permutation-invariant vectors. Clearly, this implies that
the period matrix is invariant under permutations, \textit{ie}  $\gamma_{\sigma(j)\sigma(k)}=\gamma_{jk}$. It thus follows that there
exist integers $ \beta_j \in \mathbb{Z}$ and a non-trivial permutation $\pi \in \mathfrak{S}_M$ such that
\beq
I_p=I_{\pi(p)} + (a-b) \beta_p  + b \sum_{\ell=1}^{M} \beta_{\ell} \; .
\label{equation I et I permute}
\eeq
Summing \eqref{equation I et I permute} over all the $p$'s we get that
\beq
   \bigl(a+(M-1)b \bigr) \sum_{\ell=1}^{M} \beta_{\ell}= 0\quad \Rightarrow\quad \sum_{\ell=1}^{M} \beta_{\ell}=0,
\eeq
the latter stemming from the fact that the period matrix $G$ is assumed to have a non-vanishing determinant, \textit{cf} \eqref{eq:numG}. Thus, we get an expression for $\beta_p$ in terms of the parameters $t_p$ describing the points $I_p$, namely
\beq
\beta_p=t_p-t_{\pi(p)} \in(-1,1).
\eeq
As $\beta_p \in \mathbb{Z}$, we necessarily have $\beta_p=0$ for all $p$. As ${\bf I}$ is an off-diagonal element
$(I_1,\dots,I_M)=(I_{\pi(1)},\dots,I_{\pi(M)})$ implies that $\pi=\id$, which contradicts the fact that $\sigma \not= \sigma^{\prime}$. \endproof

\vspace{3mm}

The above lemma leads to a complete description of the solutions of the generalized Bethe equations in the trigonometric case.
\begin{prop}
\label{proposition correspondance 1-a-1 avec cas trig et factor module}
The solutions of the trigonometric generalized Bethe equations \Ref{bethe} are in a one-to-one correspondence
with the regular orbits of the factor-module 
\be
  \mathbb{G}\equiv\mathbb{Z}^M/ \left<G\right>
\ee
with respect to the action of permutation group $\mathfrak{S}_M$.
\end{prop}
In the remaining part of this Section, we will prove
\begin{theo}
\label{theorem nombre element factor module}
For a matrix $G$ of the form \eqref{def-g}, where $a$ and $b$
are not necessarily given by \eqref{eq:def-ab},
the number of regular orbits of the factor module $\mathbb{G}=\mathbb{Z}^M/\left<G\right>$ is
\beq\label{eq:cmab-hypothesis}
    C_M^{[a,b]}=\frac{\abs{a+(M-1)b}}{M}\, \binom{\abs{a-b}-1}{M-1}.               
\eeq
\end{theo}

Taking $a=N+(M-1)K$ and $b=-K$ we get that
\begin{cor}
The number of solutions to the generalized Bethe ansatz equation in the trigonometric case is given, for $N=1$,
by the Fuss-Catalan numbers $C_M^{(K,1)}$, and by  
\beq
    C_M^{(K,N)}=\frac{N}{M}\,\binom{KM+N-1}{M-1}=\frac{N}{KM+N}\,\binom{KM+N}{M},
%\label{eq:cmkn-hypothesis}
\eeq
for arbitrary $N$. Note that if $C(x)=\sum_{M=0}^{+\infty} C_{M}^{(K,1)} t^M$ is the generating function of the Fuss-Catalan numbers then $C^{N}(x) =\sum_{M=0}^{+\infty} C_{M}^{(K,N)} t^M$ is the generating function of the numbers $C_M^{(K,N)}$.

\end{cor}

\subsection{Number of elements in $\mathbb{G}$}

%\subsubsection{A few properties of integer matrices}

\begin{prop}
\label{proposition cardinal de factor module}
Let $\mathbb{G}$ be a factor module of $\mathbb{Z}^M$ defined by some period matrix $G\in \matmz$ such that $\det G\neq0$,
then  $\mathbb{G}$ is a finite set, and its cardinality,
up to an eventual sign, equals to the determinant of $G$:
\beq
    \#\mathbb{G}=\abs{\det G}.
\eeq
\end{prop}

\noindent{\bf Proof.} It follows from elementary algebra that the matrix $G$ admits a Smith normal form:
\beq\label{eq:smith}
     G=UDV,\qquad U, D, V\in \matmz \; , \; D=\text{\rm diag}(d_1,\ldots,d_M) \;.
\eeq
and the matrices $U$ and $V$ are invertible (\textit{ie} $\det U = \pm 1 \, , \; \det V = \pm 1$).

Substituting \Ref{eq:smith} into the set of relations \Ref{eq:Ge=0} determining $\mathbb{G}$ we obtain
\beq
    UDV\boldsymbol{e}=0 \; .
\eeq

The matrix $U$ being invertible, we multiply the relations by $U^{-1}$ from the left
and obtain an equivalent system of relations.
On the other hand,  due to the invertibility of $V$,
the system of generators $\boldsymbol{e}$ for the module $\Zbbd^M$ can be replaced by
the equivalent system of generators $\boldsymbol{e'}=V\boldsymbol{e}$.

As a result, we get an equivalent description of the factor-module
$\mathbb{G}$ in terms of the generators $\boldsymbol{e'}$ and the relations
\beq
       D\boldsymbol{e'}=0 \; , \qquad  ie \quad
      d_me_m^\prime=0,\quad  \text{for} \quad m=1,\ldots,M \; .
\label{eq:Dmmem}
\eeq

Thus, the factor-module $\mathbb{G}$ decomposes into a direct sum of rank 1 modules,
each with the single generator $e_m^\prime$ and the single relation $d_me_m^\prime=0$.
Obviously, each component consists of $\abs{d_m}$ elements
that is equivalence classes whose representatives can be chosen,
for example, as
\beq
    0,\,e_m,\,2e_m,\,3e_m,\ldots,(\abs{d_m}-1)e_m \; .
\eeq

Thus the cardinality of $\mathbb{G}$ equals $\prod_m \abs{d_m}=\abs{\det D}$.
It remains to note that, by virtue of \Ref{eq:smith} $\abs{\det G}=\abs{\det D}$.
\endproof

\begin{cor}
Let $\mathbb{G}$ be the factor module defined by the period matrix $G$ \eqref{def-g}. Then
\beq\label{eq:numG}
    \#\mathbb{G}=\abs{\det G}=\abs{\bigl(a+(M-1)b\bigr)(a-b)^{M-1}} \;.
\eeq
\end{cor}

\noindent{\bf Proof. }
In the case of the matrix $G$ of interest \eqref{def-g}, $\det G$ can be computed thanks to the well-known
formula for the determinant of rank-1 perturbation of a matrix. Let $A$ be an arbitrary square matrix, $x$ be a vector-column, and $y^t$ be a vector-row. Then
\beq\label{eq:proj-pert}
    \det(A+xy^t)=\det A+y^t \check{A} x,
\eeq
where $\check{A}$ is the adjoint matrix that is $A\check{A}=\det A$.

\vskip2mm
In the case of interest we have $G=A+xy^t$, where
\beq
   A=(a-b)\,\text{id}_M,\qquad
   x=\begin{pmatrix}
      1\\\vdots\\1
   \end{pmatrix},\qquad
   y^t=(b,\ldots,b).
\eeq
\endproof

\vspace{3mm}

As we have already pointed out,  formula \Ref{eq:numG}, constitutes only the first step towards the
answer; what we need to count are not all the points of $\mathbb{G}$
but the regular orbits in respect to the action of the group $\mathfrak{S}_M$.

%%%%%%%%%%%%%%%%%%%%%%%%%%%%%%%%%%%%%%%%%%%%%%%%%%%%%%%%%%%%%%%%%%%%%%%%%%%%%%%%
\subsection{The diagonal submodules of $\mathbb{G}$}

In order to describe the diagonal submodules of $\mathbb{G}$, we start by presenting an example.

\vspace{2mm}
\noindent{\bf Example.} When $M=2$, the period matrix
\beq
    G=\begin{pmatrix} a & b\\ b & a\end{pmatrix}
\eeq
produces the system of relations
\beq\label{eq:M2ab-rels}
    \left\{\begin{array}{r}
       ae_1+be_2=0\; ,\\
       be_1+ae_2=0\; .
    \end{array}\right.
\eeq

Diagonals correspond to $I_1=I_2\equiv I_{12}$ that is to the points
$I_{12}(e_1+e_2)$.
In other words, the element $e_1+e_2$ generates a submodule, which
we shall call a {\em diagonal} submodule.
The relation for this submodule is obtained by adding together
the equations \Ref{eq:M2ab-rels}:
\beq
    (a+b)(e_1+e_2)=0.
\eeq
The size of the submodule (the number of points on the diagonal)
is obviously $\abs{a+b}$.
\vspace{2mm}

Let $\mathcal{P}$ be a partition of the set $\{1,\ldots,M\}$
into $J$ disjoint non-empty subsets $\mathcal{P}=\{\mathcal{P}_1,\, \ldots\, ,\mathcal{P}_J\}$.
To each partition $\mathcal{P}$ we associate the diagonal submodule $\mathcal{D}_{\mathcal{P}}$
generated by $J$ generators $e_{\mathcal{P}_j}\equiv\sum_{i\in \mathcal{P}_j} e_i$.
Its matrix of relations is obtained from the matrix $G$ by adding the rows labeled by
$i\in \mathcal{P}_j$.

\vspace{2mm}
\noindent{\bf Example. }
For $M=3$ there are 5 partitions of the set $\{1,2,3\}$:
the 3-component one: $\{\{1\},\{2\},\{3\}\}$ corresponding to the original module
$\left<G\right>$, and 4 other partitions generating diagonal modules:
three 2-component ones:
$\{\{1,2\},\{3\}\}$, $\{\{1,3\},\{2\}\}$, $\{\{2,3\},\{1\}\}$
and one 1-component: $\{\{1,2,3\}\}$.

The 2-component partition $\{\{1,2\},\{3\}\}$ gives rise to the generators
$\{e_1+e_2\equiv e_{12},e_3\}$ and relations $(a+b)e_{12}+2be_3=be_{12}+ae_3=0$,
similarly for the rest two. The 1-component partition
$\{\{1,2,3\}\}$ gives rise to a single generator $e_1+e_2+e_3\equiv e_{123}$
and a single relation $(a+2b)e_{123}$ .

\vspace{2mm}

\begin{prop}
Let $\mathcal{P}=\{\mathcal{P}_1,\, \ldots\, ,\mathcal{P}_J\}$ be a partition of $\{1,\ldots,M\}$.
Then the associated diagonal submodule $D_{\mathcal{P}}$ is isomorphic to the diagonal submodule $\mathcal{D}_{\lambda}$ corresponding to the partition $\mathcal{P}_1=\{1,\dots,\la_1\}$, $\dots$, $\mathcal{P}_J=\{M-\la_J+1,\dots, M\}$, where $\la_j=\# \mathcal{P}_j$ gives the number of elements
in the partition $\mathcal{P}_j$.
\end{prop}

\noindent{\bf Proof.}
Taking the cardinalities $\#\mathcal{P}_j$ of the components of the partition $\mathcal{P}$
of the {\em set} $\{1,\ldots,M\}$ and rearranging them in the decreasing order
we obtain a partition
$\la=\{\la_1\geq\la_2\geq\ldots\geq\la_J\}$ of the {\em number} $M$ into
$J$ nonzero numbers: $\abs{\la}\equiv\la_1+\ldots+\la_J=M$.
Clearly, partitions $\mathcal{P}$ corresponding to the same partition
$\la$ are related by permutations of the numbers
$\{1,\ldots,M\}$, and the corresponding diagonal submodules
are isomorphic. \endproof

\vspace{3mm}

\begin{cor}\label{corollaire nombre element diag module}
The number of elements in every diagonal submodule $\mathcal{D}_{\mathcal{P}}$ only depends on the number $J$ of components of the partition $\mathcal{P}=\{\mathcal{P}_1,\, \ldots\, ,\mathcal{P}_J\}$ and is given by
\beq\label{eq:detGla}
   \# \mathcal{D}_{\mathcal{P}} \equiv  \D^M_J =\bigl(a+(M-1)b\bigr)\bigl(a-b\bigr)^{J-1}.
\eeq
\end{cor}

\noindent{\bf Proof.}
As $\mathcal{D}_{\mathcal{P}}$ and $\mathcal{D}_{\lambda}$ are isomorphic, is is enough to count the number of elements if $\mathcal{D}_{\lambda}$. The matrix elements of period matrix $G_\la$ for the diagonal submodule $\mathcal{D}_{\lambda}$ are given by
\beq\label{eq:gjkla}
   g_{jk}^\la=\d_{jk}\,(\bigl(a+(\la_j-1)b\bigr)+(1-\d_{jk})\,\la_j b, \qquad
   j,k=1,\ldots,J.
\eeq
Rewriting \Ref{eq:gjkla} as
\beq
   g_{jk}^\la=(a-b)\,\d_{jk}+\la_j b,
\eeq
we notice that $G_\la$ decomposes into a sum of a diagonal matrix
$A=(a-b)\,\text{id}_J$
and a rank 1 matrix $xy^t$, where $x_j=\la_j$ and $y_k=b$.
Applying equation \eqref{eq:proj-pert} we obtain the claim. \endproof

%%%%%%%%%%%%%%%%%%%%%%%%%%%%%%%%%%%%%%%%%%%%%%%%%%%%%%%%%%%%%%%%%%%%%%%%%%%%%%%%

\subsection{Elimination of diagonals and Stirling numbers: proof of theorem \ref{theorem nombre element factor module}}

We are now in position to prove Theorem \ref{theorem nombre element factor module}.

The case $M=2$ was analyzed at the beginning of the previous Section.
There, one has the decomposition
\beq
  2!\,C_2^{[a,b]}= \abs{\det G_{11}}-\abs{\det G_2}
  =\abs{\D^2_2}-\abs{\D^2_1}
  =\abs{a+b}(\abs{a-b}-1).
\eeq

For $M=3$ we have 3 partitions of the set $\{1,2,3\}$ of the type $\la=(21)$.
Subtracting the corresponding points count from the number of points in the
original module $\mathbb{G}$ we obtain
\be
   \abs{\det G_{111}}-3\abs{\det G_{21}}.
\ee

Note, however, that each of the 3 diagonals $I_1=I_2$, $I_1=I_3$, $I_2=I_3$
contains the common sub-diagonal $I_1=I_2=I_3$. We have thus subtracted it thrice.
To compensate, we need to add it back twice. Finally, we get:
\begin{align}
    3!\,C_3^{[a,b]}&=\abs{\det G_{111}}-3\abs{\det G_{21}}+2\abs{\det G_3}\notag\\
    &=\abs{\D^3_3}-3\abs{\D^3_2}+2\abs{\D^3_1}\notag\\
    &=\abs{a+2b}(\abs{a-b}-1)(\abs{a-b}-2).
\end{align}

Proceeding similarly for $M>3$ and using corollary \ref{corollaire nombre element diag module}
stating that $\det G_{\mathcal{P}}$ only depends on $J$
we arrive to the following expansion:
\begin{align}\label{eq:CMab-expand}
    M!\,C_M^{[a,b]}&=\sum_{J=1}^M (-1)^{M-J} s_{MJ} \abs{\D^M_J}\notag\\
    &=\abs{a+(M-1)b}\sum_{J=1}^M (-1)^{M-J} s_{MJ} \abs{(a-b)^{J-1}},
\end{align}
where $s_{MJ}$ are some combinatorial coefficients that {\em do not depend} on
$a$ and $b$.

To determine $s_{MJ}$ we invoke their independence on $a$ and $b$
and set $b=0$. Then, the factor-module $\mathbb{G}$ becomes a hypercube having sides of length
$\abs{a}$. The regular orbits are labeled by the sequences
$0\leq I_1<I_2<\ldots<I_M\leq \abs{a}-1$. The number of such points
is obviously
\beq\label{eq:CMa0}
    C_M^{[a,0]}=\binom{\abs{a}}{M}=\frac{\abs{a}(\abs{a}-1)\ldots(\abs{a}-M+1)}{M!}.
\eeq

Setting $b=0$ in \Ref{eq:CMab-expand} and substituting \Ref{eq:CMa0},
we obtain its expansion into powers of $\abs{a}$
\beq
  \abs{a}(\abs{a}-1)\ldots(\abs{a}-M+1)
   =\sum_{J=1}^M (-1)^{M-J} s_{MJ} \abs{a}^J \;.
\eeq
The latter coincides with the definition of the  (unsigned)
{\em Stirling numbers of 1st kind} \cite{GrahamKnuthPatashnikConcreteMathematics,StanleyEnumerativeCombinatorics}.
In Knuth's notation \cite{GrahamKnuthPatashnikConcreteMathematics}:
\beq
       s_{MJ}=\stirling{M}{J}.
\eeq

Returning to the general case \Ref{eq:CMab-expand}
and using the definition of the Stirling numbers we obtain
\begin{align*}
  M!\,C_M^{[a,b]}&=\abs{a+(M-1)b}\sum_{J=1}^M (-1)^{M-J} \stirling{M}{J} \abs{(a-b)^{J-1}}\\
    &= \frac{\abs{a+(M-1)b}}{\abs{(a-b)}}\, \abs{a-b}(\abs{a-b}-1)\ldots(\abs{a-b}-M+1)\\
  &=\abs{a+(M-1)b}\, (\abs{a-b}-1)\ldots(\abs{a-b}-M+1).
\end{align*}

The last expression coincides with \Ref{eq:cmab-hypothesis}. \endproof

%%%%%%%%%%%%%%%%%%%%%%%%%%%%%%%%%%%%%%%%%%%%%%%%%%%%%%%%

\section{Proof of theorem \ref{theorem integers with constraints}}
\label{section preuve theroem comptage cas rationel}

In this Section we prove theorem \ref{theorem integers with constraints}. The proof is carried out by establishing a
bijection between the solutions to the generalized Bethe equation in the trigonometric and
rational cases.

One might expect that the solutions to the generalized Bethe equations in the  trigonometric case  with parameters $\{ \eps \eta_k \}$ and $\{ \eps \xi_k \}$, scale as $\text{O}(\eps)$ in the $\eps\rightarrow0^+$ limit (up to some possible shifts by $ \left(\pi \mathbb{Z}\right)^M$). If this is indeed the case, then the rational equations can be considered as a degeneration of the trigonometric ones. Of course, one should be rather careful in such a limiting procedure. When $\eps\rightarrow0^+$, it could be that some solutions to the trigonometric equations may not go to zero, and thus one might loose some solutions in the procedure. In particular, there might not be a one-to-one correspondence between the two sets of solutions. We prove that this does not happen:
one can build, for $\eps$ small enough, a one-to-one correspondence between the solutions to the generalized Bethe equations in the trigonometric and rational cases.

From now on, we assume that the parameters $\{ \xi_k \}$ and $\{ \eta_k\}$ occurring in the trigonometric case
are rescaled by $\eps>0$. Therefore, the trigonometric functions $F$ and $S$ are given by
\begin{align*}
F(u) &= F_{\infty} \prod_{n=1}^{N} \frac{ \sin(u+ \eps \xi_n) }{ \sin(u+ \eps \overline{\xi}_n) }  \qquad
\text{where} \;\;  F_{\infty} = \text{e}^{-2i\pi \varphi_{\infty}} \; , \;\;\;  -1 < (M+1) \varphi_{\infty} <0 \\
S(u) &=  \prod_{n=1}^{K} \frac{ \sin(u+\eps \eta_n) }{ \sin(u+ \eps \overline{\eta}_n) } \; .
\end{align*}
In the following, the parameters $\{\eta_k\}_{k=1}^{K}$ and $\{\xi_r\}_{r=1}^{N}$ are to be considered fixed. The only varying parameter will be $\eps$. We will call it the dilatation parameter.
From now on, we shall \textit{only focus} on the solutions ${\boldsymbol{v}}$ to the trigonometric GBE
belonging to the fundamental domain $(-\pi/2, \pi/2]$, \textit{ie} for all $j \in \{1,\dots,M\}$,
$v_j \in (-\pi/2, \pi/2]$.
We also introduce the following definitions for the function $\varphi_{\eps}$ and $\theta_{\eps}$:
\begin{align*}
    \varphi_{\eps}(u) &=  2\pi\varphi_\infty
    +\sum_{n=1}^N \chi\bigl(u+\eps\Re(\xi_n),\eps\Im(\xi_n)\bigr),\\
    \theta_{\eps}(u) &=
    \sum_{j=1}^K \chi\bigl(u+\eps\Re(\eta_j),\eps\Im(\eta_j)\bigr)\; ,
\end{align*}%
where $\chi(u,c)$ is given by \eqref{definition fonction chi}.

%%%%%%%%%%%%%%%%%%%%%%%%%%%%%%%%%%%%%%%%%%%%%%%%%%%%%%%%%%%%%%%%
\subsection{Rational constraints on the trigonometric integers}

In this section we prove the following theorem

\begin{theo}
\label{theorem cas trig a les contraintes rat}
There exists $\eps_0$ small enough such that any sequence $I_1,\dots,I_M$  defined by  a solution
to the trigonometric GBAE \eqref{logbethe} with dilatation parameter $\eps \in (0,\eps_0)$
satisfies the inequalities:
\begin{equation}
N r -1  -K \frac{r(r+1)}{2} +K M r  \geq \sum_{k\in \mathcal{J} }^{} I_k \geq K \frac{r(r-1)}{2}  \quad ,
 \qquad \forall \mathcal{J} \subset \{1,\dots,M \}.
\label{inegalite entiers bis}
\end{equation}
\end{theo}

The main consequence of this theorem is that any solutions of the trigonometric
GBE with sufficiently small inhomogeneities $\{\eps \eta_k\}$, gives rise to a set of $M$ integers $(I_1,\dots,I_M)$
satisfying the set of inequalities issued from the minimum condition of the rational potential.
 This means that there is at most as much solutions to the trigonometric GBE as there are solutions to the rational GBE.

To prove the theorem we need to establish two preparatory lemmas. We first observe that for $\eps$ small enough, any coordinates $v_j$ of a solution $\boldsymbol{v}$ to the trigonometric logarithmic Bethe equations stays uniformly away from the boundaries $\pm \pi/2$.

\begin{lemme}
\label{proposition Confinement Bethe Roots}
There exists an $\eps_0>0$ and an $\alpha>0$ such that  given any solution $\boldsymbol{v}=(v_1,\dots,v_M)$ to the trigonometric logarithmic
GBE \eqref{logbethe} with dilatation parameter $\eps \in (0,\eps_0)$, one has
\begin{equation}
v_j \in \left[-\frac{\pi}{2}+\alpha, \frac{\pi}{2}-\alpha\right] \quad \text{for} \quad  j=1,\dots,M \; .
\end{equation}
\end{lemme}

\noindent{\bf Proof.}
We prove the statement by contradiction. Assume that there exists a sequence $\eps_n \rightarrow 0$
and a sequence ${\boldsymbol{v}}^{(n)}$ of solutions to the logarithmic GBE arising in the trigonometric case and corresponding to a dilatation parameter $\eps_n$  such that
\begin{equation}
v_1^{(n)} \underset{n \rightarrow  +\infty}{\longrightarrow} -\frac{\pi}{2}  
\quad  \text{or} \quad v_M^{(n)} \underset{n \rightarrow +\infty}{\longrightarrow} \frac{\pi}{2}
\label{condition convergence suite vn}
\end{equation}

Here, we choose to work with solutions having ordered coordinates  $-\pi/2\leq v_1^{(n)}<\dots<v_M^{(n)}\leq \pi/2$.
As the sequence $\boldsymbol{v}^{(n)}$ is bounded in norm,  it admits a converging subsequence. We continue to denote this converging subsequence by  $\boldsymbol{v}^{(n)}$. It fulfills
\begin{equation}
\begin{array}{cc c l} v_s^{(n)} &\underset{n \rightarrow +\infty}{\longrightarrow } &-\pi/2 , & s \in \{1, \dots \ell \} \\
v_s^{(n)} &\underset{n \rightarrow  +\infty}{\longrightarrow} &v_s^{(\infty)} 
  \in (-\pi/2, \pi/2), & s \in \{\ell+1, \dots M-\ell^{\prime} \} \;\\
v_s^{(n)}& \underset{n \rightarrow  +\infty}{\longrightarrow} &\pi/2 , & s \in \{M-\ell^{\prime}+1, \dots, M \} \; .
\end{array}
\end{equation}
Due to \eqref{condition convergence suite vn}, at least one of the $v_j^{(n)}$ has to converge to one of the boundaries points $\pm \pi/2$. Hence, $\ell+\ell^{\prime}\geq 1$.  Taking the product of the trigonometric  GBE for $v_s^{(n)}$ where
$s$ runs through the set $\mathcal{S}= \{ 1, \ell\}\cup \{M-\ell^{\prime}+1, \dots, M \}  $ yields
\beq
1= \prod_{s\in S}^{} F(v_s^{(n)}) \cdot \prod_{s\in S}^{} \prod_{\substack{k=1 \\ k\not=s}}^{M}
S(v_s^{(n)}-v_k^{(n)})= \prod_{s\in S}^{} F(v_s^{(n)}) \cdot \prod_{s\in S}^{} \prod_{\substack{k=1 \\ k\not\in S}}^{M}
S(v_s^{(n)}-v_k^{(n)}) \; .
\label{produit equation de Bethe pour vs convergeant au bord}
\eeq

In the last equality we made use of the space parity condition satisfied by $S$ so as to cancel out the products
$S(v_s^{(n)}-v_k^{(n)}) S(v_k^{(n)}-v_s^{(n)})$ for $k,s \in \mathcal{S}$.  It is easy to see that for $s\in \mathcal{S}$, $F(v_s^{(n)}) \rightarrow  F_{\infty}$. As for $k \not\in \mathcal{S}$, $v_k^{(n)}$ converges to some point in $(-\pi/2, \pi/2)$. We have,
\begin{equation}
v_s^{(n)}-v_k^{(n)} \underset{n \rightarrow  +\infty}{\longrightarrow}
 \delta v_{sk} \in (-\pi, 0) \cup (0,\pi)  
 \qquad \forall s \in \mathcal{S} \;\; \text{and}\;\;  k\in \{1,\dots,M\}\setminus \mathcal{S} \;.
\end{equation}

In its turn, this means that $S(v_s^{(n)}-v_k^{(n)}) \rightarrow  1$.
Thus, the $n \rightarrow +\infty$ limit of \eqref{produit equation de Bethe pour vs convergeant au bord} yields
\begin{equation}
F_{\infty}^{\ell+\ell^{\prime}}=\re^{-2i\pi(\ell+\ell^{\prime})\varphi_{\infty}}=1 \; .
\end{equation}
As $1\leq\ell+\ell^{\prime}<M+1$, this contradicts the fact that $-1< (M+1)\varphi_{\infty}<0$. \endproof

\vspace{3mm}
We now observe that, if we take any two points $v, w$ lying in $(-\pi/2+\alpha, \pi/2-\alpha)$ for some $\alpha>0$,
then $\theta_{\eps}$ will be almost contained in the interval $[0, 2\pi]$, \textit{ie} $\theta_{\eps}(v-w) \in [-\nu, 2\pi+\nu]$, where $\nu\rightarrow  0$ as $\eps \rightarrow  0$.

\begin{lemme}
\label{lemme Decroissance de theta en eps}
Let $0<\alpha< \pi/4$, then there exists $\eps_0>0$ such that for all $\eps \in (0, \eps_0)$
\beq
2\pi K - 2\pi \frac{\varphi_{\infty} }{M^2} > \theta_{\eps}(v-w) >  2\pi \frac{\varphi_{\infty} }{M^2}  \qquad \qquad
\forall \; v, w \, \in \left[-\frac{\pi}{2}+\alpha, \frac{\pi}{2}-\alpha\right] \; .
\eeq
\end{lemme}

Here, we remind that $-1<\varphi_{\infty}<0$. 

\noindent{\bf Proof.}
Let $\Im\eta>0$. Then,
as $\chi$ is strictly increasing on the real axis, given $v, w \, \in [-\pi/2+\alpha, \pi/2-\alpha] $ we have
\begin{equation*}
\chi(-\pi+2\a + \eps\Re{\eta},\eps \Im{\eta})  \; \leq \;  \chi(v-w + \eps\Re{\eta},\eps \Im{\eta}) \; \leq \; \chi(\pi-2\a + \eps\Re{\eta},\eps \Im{\eta})  \;.
\end{equation*}
In the $\eps \rightarrow  0^+$ limit, the $lhs$ goes to zero whereas the $rhs$ goes to $2\pi$. Indeed, taking $\eps$
small enough so that $0<\alpha-\eps |\Re(\eta)|$, one gets
\begin{multline*}
\biggl|\chi\bigl(-\pi+2\alpha+ \eps\Re(\eta),\eps \Im(\eta) \bigr) \biggr|=
\abs{\int\limits_{-\frac{\pi}{2} }^{-\pi +2\alpha+ \eps\Re(\eta)} \!\!\! \frac{ \sinh(2 \eps \Im(\eta))  \;  }
{\sin(u+i\eps \Im(\eta)  ) \sin(u-i\eps \Im(\eta) )}  \rd u }  \\
\leq
\frac{  \sinh(2 \eps \Im(\eta)  ) \cdot \pi/2 }{  \sin(\alpha+i\eps \Im(\eta)  ) \sin(\alpha-i\eps \Im(\eta)  )  }
\underset{\eps \rightarrow  0^+}{\longrightarrow} 0 \;.
\end{multline*}
Similarly, using the quasi-periodicity of $\chi$ we get
\begin{multline*}
\biggl|\chi\bigl(\pi-2\alpha+ \eps\Re(\eta),\eps \Im(\eta) \bigr) - 2\pi \biggr| \\
=\int\limits_{-\frac{\pi}{2} }^{-2\alpha+ \eps\Re(\eta)} \!\!\! \frac{ \sinh(2 \eps \Im(\eta))\;   }
{ \sin(u+i\eps \Im(\eta)  ) \sin(u-i\eps \Im(\eta) ) } \rd u   \; \underset{\eps \rightarrow  0^+}{\longrightarrow} 0 .
\end{multline*}
It thus follows that there exists an $\eps_0>0$ such that for all
\beq
k\in \{1,\dots,K \} \;  \; \text{and}
 \;  v, w \, \in \left[-\frac{\pi}{2}+\alpha, \frac{\pi}{2}-\alpha\right]
\eeq
we have
\begin{equation}
\frac{2\pi \varphi_{\infty}}{K M^2}   < \chi\bigl(v-w+ \eps\Re(\eta_k),\eps \Im(\eta_k)\bigr)
< 2\pi - \frac{2\pi \varphi_{\infty}}{K M^2}   \;.
\end{equation}
uniformly in $\eps \in (0, \eps_0)$. Summing these inequalities over $k$ we get the claim.  \endproof

We are now in position to prove the main result of this subsection:

\vspace{2mm}
\noindent{\bf Proof} [\textit{Theorem} \ref{theorem cas trig a les contraintes rat}].
We choose an $\eps_0>0$  such that the conclusion of lemmas \ref{proposition Confinement Bethe Roots} and
\ref{lemme Decroissance de theta en eps} hold simultaneously. Let $\eps$ be such that $\eps_0>\eps>0$ and $\boldsymbol{v}$ be any solution of the trigonometric  logarithmic GBE with parameters $\{\eps \eta_k\}$ and $\{\eps \xi_p \}$. Then, according to the conclusions of lemma \ref{proposition Confinement Bethe Roots} there exists an $\alpha$ such that
$v_j \in (-\pi/2+\alpha, \pi/2-\alpha)$. By virtue of  lemma \ref{lemme Decroissance de theta en eps},
we have
\begin{equation}
     2\pi \frac{\varphi_{\infty}}{M^2} < \theta_{\eps}(v_j-v_k) < 2\pi K -2\pi \frac{\varphi_{\infty}}{M^2} \;.
\end{equation}

Also, since $v_j\in (-\pi/2+\alpha, \pi/2-\alpha)$ and one can always assume $\eps$ to be small enough so that $\alpha-\eps|\Re(\xi_k)|>0$ for $k=1,\ldots,N$, we have
\begin{equation}
2\pi(N+\varphi_{\infty})> \varphi_{\eps}(v_j)> 2\pi \varphi_{\infty} \;.
\end{equation}
\noindent Let $\mathcal{J} \subset \{1,\dots,M\}$ such that $\# \mathcal{J} =r$, then
\begin{align*}
2\pi \sum_{k\in \mathcal{J} }^{} I_k&= \sum_{k\in \mathcal{J}}^{} \varphi_{\eps}(v_k) +
\sum_{k\in \mathcal{J} }^{} \sum_{j\not=k}{} \theta_{\eps}(v_k -v_j) \\
&= \sum_{ \substack{j,k\in \mathcal{J} \\ j>k} }^{} 
  \bigl[\theta_{\eps}(v_k -v_j)+\theta_{\eps}(v_j -v_k)\bigr]+
\sum_{k\in \mathcal{J} }^{} \varphi_{\eps}(v_k) + \sum_{k\in \mathcal{J} }^{}\sum_{ \substack{j=1 \\ j\not\in \mathcal{J}} }^{M}
\theta_{\eps}(v_k -v_j) \\
&\geq  2\pi K \frac{r(r-1)}{2} +2\pi r \varphi_{\infty} +2\pi \frac{r(M-r)}{M^2}\,\varphi_{\infty} \\
&\geq 2\pi (r+1) \varphi_{\infty} + 2\pi K \frac{r(r-1)}{2}\; .
\end{align*}

It remains to use the bound $-1< (r+1) \varphi_{\infty}<0$ and invoke the fact that $\sum_{k\in S}^{} I_k$ is an integer so as to obtain the equality in the $rhs$ of \eqref{inegalite entiers bis}. 

One gets the second set of inequalities in a similar way
\begin{align*}
2\pi \sum_{k\in \mathcal{J} }^{} I_k
&= \sum_{ \substack{j,k\in \mathcal{J} \\ j>k} }^{} 
  \bigl[\theta_{\eps}( v_j -v_k )+\theta_{\eps}( v_k -v_j )\bigr] +
\sum_{k \in  \mathcal{J}}^{} \varphi(v_k) + \sum_{k\in \mathcal{J}}^{} \sum_{j\not\in \mathcal{J}}{} \theta_{\eps}( v_k -v_j ) \\
&< 2\pi K \frac{r(r-1)}{2} + r 2\pi \varphi_{\infty} +2\pi N r +  2\pi K (M-r) r
 - 2\pi \frac{r(M-r)}{M^2}\, \varphi_{\infty}  \\
&\leq 2\pi \left( N r -1  -K \frac{r(r+1)}{2} +K M r \right) \; .
\end{align*}

When passing to the last line, we again have used the fact that $\sum_{k\in \mathcal{J}}^{} I_k$ is an integer.  \endproof

%%%%%%%%%%%%%%%%%%%%%%%%%%%%%%%%%%%%%%%%%%%%%%%%%%%%%%%%%%%%%%%%%%%%%%%%%%%%%%
\subsection{From the rational to the trigonometric solutions}

One can visualize the generalized Bethe equation as the set of constraints defining the real valued zeroes
${\boldsymbol{v}}$ with pairwise distinct coordinates of the mapping
\begin{align*}
\mathcal{Y}_{\{\eta\},\{\xi\}}: \Cbbd^M & \rightarrow \Cbbd^M \\
\phantom{\mathcal{Y}:} {\bf z} & \mapsto
\left(
\mathcal{Q}_{\{ \eta\},\{ \xi\}}(z_1 \mid {\bf z} ), \dots , \mathcal{Q}_{\{ \eta\},\{ \xi\}}(z_M \mid {\bf z} ) %
\right).
\label{definition fonction Y}
\end{align*}

We remind that the trigonometric Baxter polynomial $\mathcal{Q}_{\{ \eta\},\{ \xi\}}(z_M \mid {\bf z} )$ has been introduced in  \eqref{definition Q}.

\begin{prop}
Let ${\boldsymbol{v}} $ be a solution of the rational generalized Bethe equations with parameters $\{ \eta_k \}$ and $\{ \xi_k \}$. Then, there exists an $\eps_{\boldsymbol{v} }>0$
and a mapping $g_{ \boldsymbol{v} }(\eps)$ from the interval $(-\eps_{\boldsymbol{v}}, \eps_{\boldsymbol{v}})$ to some open neighborhood of $0$ in $\Rbbd^M$ such that $\eps g_{ \boldsymbol{v}}(\eps)$ is a solution of the trigonometric generalized Bethe equations with parameters
$\{\eps \eta\}$ and $\{\eps \xi\}$.
\end{prop}

\noindent{\bf Proof.}
The function $\mathcal{G}$ given by
\begin{align*}
\mathcal{G} : (-1;1) \times \Rbbd^M  & \rightarrow  \Cbbd^M \\
\phantom{\mathcal{G} :} (\eps, \bf z) & \mapsto i \eps^{-N-MK} 
\mathcal{Y}_{ \{\eps \eta \},\{\eps \xi\} } (\bf \eps \,  z)
\end{align*}
is continuously differentiable and
\beq
\lim_{ \substack{ \eps \rightarrow 0 \\ {\bf z}  \rightarrow {\boldsymbol{v} }}  } \frac{\partial}{\partial z_j} \cdot \mathcal{G}_k (\eps, {\bf z})=F_{\infty}
\left\{ \prod_{r=1}^{N} (v_k+\xi_r) \cdot
\prod_{m=1}^{M} \prod_{\ell=1}^{K} (v_{k}-v_m+\eta_{\ell})  \right\}
\frac{\partial^2}{\partial z_j \partial z_k} \cdot W({\bf z}) \mid_{ {\bf z}={\boldsymbol{v}}  } \; .
\eeq
Here $W$ is the strictly convex potential \eqref{definition potential W} appearing in the rational case. As any solution to the generalized Bethe equations is real,
the product in the pre-factor in the $rhs$ is non-zero. We have shown in proposition \ref{proposition unicite des solutions} that $W$ has a positive defined Hessian
\beq
\det\left[ \frac{\partial^2}{\partial z_j \partial z_k} \cdot W({\bf z})\right] _{\mid_{ {\bf z}={\boldsymbol{v}}  }  } >0,
\eeq
therefore  $\det_M \left[ \partial_{z_k}\mathcal{G}_{\ell} (0,\boldsymbol{v}) \right] \not=0 $ and we can apply the implicit function theorem to the mapping $\mathcal{G}$ in the vicinity of the point $(0,\boldsymbol{v})$.

There exists open neighborhoods  $(-\eps_{\boldsymbol{v}}, \eps_{\boldsymbol{v}})$ of $0$, and 
$V_{\boldsymbol{v}}$ of $\boldsymbol{v}$ and a $\mathcal{C}^1$ mapping $g_{\boldsymbol{v}}:   (-\eps_{\boldsymbol{v}}, \eps_{\boldsymbol{v}})  \rightarrow  V_{\boldsymbol{v}} $, 
 such that $g_{\boldsymbol{v}}(0)= {\boldsymbol{v}}  $ and 
\beq
\left\{ (\eps,{\bf z})  \in (-\eps_{\boldsymbol{v}}, \eps_{\boldsymbol{v}}) \times V_{\boldsymbol{v}} \; : \;
                    \mathcal{Y}_{\{ \eps\eta \}, \{\eps \xi \} } ^{} (\eps \boldsymbol{z})=0  \right\}=
\left\{ \; ( \eps, g_{\boldsymbol{v}}(\eps) ): \eps  \in (-\eps_{\boldsymbol{v}}, \eps_{\boldsymbol{v}})^{\hspace{1mm}} \right\} \; .
\eeq
$g_{\boldsymbol{v}}(\eps)$ is thus a solution of the trigonometric GBE with parameters $\{\eps \eta\}$ and
$\{\eps \xi \}$.
 \endproof

\vspace{5mm}
\begin{prop}
Let $\boldsymbol{v}$ and $\boldsymbol{v}^{\prime}$ be two distinct solutions of the rational GBE.
Let $\eps_{0}= \min ( \eps_{\boldsymbol{v}}, \eps_{\boldsymbol{v}^{\prime}}) >0$, where $\eps_{\boldsymbol{v}}$
 is as given in the above proposition.
Then, for any $\eps \in (-\eps_0, \eps_0)$,
$g_{\boldsymbol{v}}(\eps) \not = g_{\boldsymbol{v}^{\prime}}(\eps)$.
\end{prop}
\noindent{\bf Proof.}
As $\eps \mapsto g_{\boldsymbol{v}}(\eps)$ is continuous, it follows that
\beq
\varphi_{\eps}\left( \eps [g_{\boldsymbol{v}}]_k(\eps)  \right) \quad \text{and} \quad
\theta_{\eps}\left( \eps [g_{\boldsymbol{v}}]_k(\eps)-\eps [g_{\boldsymbol{v}}]_k(\eps)  \right)
\eeq
are all continuous in $\eps$ on $(-\eps_0, \eps_0)$.
Hence, the combination
\begin{equation}
\eps \mapsto \varphi_{\eps}\left( \eps [g_{\boldsymbol{v}}]_k(\eps)  \right)
 + \sum_{\substack{j=1 \\ j \not= k}}^{M }
\theta_{\eps}\bigl( \eps [g_{\boldsymbol{v}}]_k(\eps)-\eps [g_{\boldsymbol{v}}]_k(\eps)  \bigr)
\label{equation GBAE continues en epsilon}
\end{equation}
is also continuous.
However, for any $\eps$ small enough, $\eps g_{\boldsymbol{v}}(\eps)$  solves the trigonometric generalized Bethe equations. Hence, the function defined in \eqref{equation GBAE continues en epsilon}
is integer valued. Being a continuous function of $\eps$, we deduce that it is constant. The value of this constant can, for instance, be determined by setting $\eps=0$. It thus follows that given two distinct solutions ${\boldsymbol{v}}$ and ${\boldsymbol{v}^{\prime}}$ of the rational GBE, one obtains two solutions of the trigonometric GBE that are characterized by two distinct
set of integers ${\bf I}$ and ${\bf I'}$. Therefore these solutions cannot coincide. \endproof

\vspace{5mm}
Hence, for $0<\eps<\eps_0$ the family $\{ \eps \, g_{\boldsymbol{v}}(\eps)\}$, where ${\boldsymbol{v}}$ solves the rational GBE,
provides as many distinct solutions of the trigonometric GBE with parameters $\{\eps \eta\}$ and $\{\eps \xi \}$
as there are solutions to the rational GBE with parameters $\{\eta\}$ and $\{\xi \}$.
We have just established that there are at least as many solutions to the trigonometric GBE as there are  to the rational ones. This ends the proof of theorem \ref{theorem integers with constraints}.

%%%%%%%%%%%%%%%%%%%%%%%%%%%%%%%%%%%%%%%%%%%%%%%%%%%%%%%%%%%%%%%%%%%%%%%%%%%%%%%%%%%%%%%%%%%%
\section{Conclusion}

In this article, we have studied a generalization of the Bethe equation where the $S$-matrix
has several poles and zeroes in the complex plane. We have characterized the set of solutions in the so-called repulsive regime. In particular, we have provided a thorough count of the number of solutions to these equations. We showed that this number is given or closely related to the Fuss-Catalan numbers. This allowed us to establish two yet-unknown combinatorial interpretations of these numbers.
On the one hand, we have shown that the Fuss-Catalan numbers count the number of integers satisfying certain
inequalities and on the other hand they give the number of regular orbits in certain factor modules.

For the moment, such generalized Bethe equation do not correspond to any integrable model. We plan to investigate
integrable models giving rise to such generalized Bethe equations in the future.
It also seems extremely appealing to analyse the combinatorial identities arising from other classes of Bethe equations then those given in \eqref{bethe}. For instance, estimating the number of solutions to the multi-pole generalizations of Gaudin-type equations should also lead to combinatorial problems for counting integers subject to sets of constraints.

%%%%%%%%%%%%%%%%%%%%%%%%%%%%%%%%%%%%%%%%%%%%%%%%%%%%%%%%%%%%%%%%%%%%%%%%%%%%%%%%%%%%%%%%
%%%%%%%%%%%%%%%%%%%%%%%%%%%%%%%%%%%%%%%%%%%%%%%%%%%%%%%%%%%%%%%%%%%%%%%%%%%%%%%%%%%%%%%%
%%%%%%%%%%%%%%%%%%%%%%%%%%%%%%%%%%%%%%%%%%%%%%%%%%%%%%%%%%%%%%%%%%%%%%%%%%%%%%%%%%%%%%%%
%%%%%%%%%%%%%%%%%%%%%%%%%%%%%%%%%%%%%%%%%%%%%%%%%%%%%%%%%%%%%%%%%%%%%%%%%%%%%%%%%%%%%%%%
%%%%%%%%%%%%%%%%%%%%%%%%%%%%%%%%%%%%%%%%%%%%%%%%%%%%%%%%%%%%%%%%%%%%%%%%%%%%%%%%%%%%%%%%
%%%%%%%%%%%%%%%%%%%%%%%%%%%%%%%%%%%%%%%%%%%%%%%%%%%%%%%%%%%%%%%%%%%%%%%%%%%%%%%%%%%%%%%%

\section*{Acknowledgements}

K. K. K. is supported by CNRS and the ANR program  ANR-10-BLAN-0120-04-DIADEMS. When part of this work was done, K.K.K. was supported by the ANR program GIMP
ANR-05-BLAN-0029-01 and then by the EU Marie-Curie Excellence Grant MEXT-CT-2006-042695. K. K. K would like to thank the Departments of Mathematics of the University of York for hospitality.

%\bibliographystyle{amsplain}
%\bibliography{bibliotemple}
\providecommand{\bysame}{\leavevmode\hbox to3em{\hrulefill}\thinspace}
\providecommand{\MR}{\relax\ifhmode\unskip\space\fi MR }
\providecommand{\MRhref}[2]{%
  \href{http://www.ams.org/mathscinet-getitem?mr=#1}{#2}
}
\providecommand{\href}[2]{#2}

\end{document}